\documentclass[12pt]{emulateapj}
\usepackage{epsfig}
\usepackage{natbib}

\newlength{\colwidth}

\newcommand{\be}{\begin{equation}}
\newcommand{\ee}{\end{equation}}
\newcommand{\bee}{\begin{eqnarray}}
\newcommand{\eee}{\end{eqnarray}}

\newcommand{\ov}{ \Omega_{\rm \Lambda} } 
\newcommand{\om}{ \Omega_{\rm m} } 

\newcommand{\mstar}{ M_{*}(a) }
\newcommand{\msol}{\hbox{${\rm M}_\odot$}} 
\newcommand{\rvir}{\hbox{$r_{\rm 200}$}} 
\newcommand{\mvir}{\hbox{$M_{\rm 200}$}} 
 
\newcommand{\mtrunc}{M_{c}}
\newcommand{\LCDM}{$\Lambda$CDM }
\newcommand{\sigate}{ \sigma_{8}}
\newcommand{\actoy}{a_{c,{\rm toy}}}

\newcommand{\hinv}{h^{-1}} 
\newcommand{\mpc}{\rm{Mpc}} 
\newcommand{\kpc}{\rm{kpc}}

\newcommand{\sig}{ \langle \sigma v \rangle }

\newcommand{\lsim}{\lower.5ex\hbox{\ltsima}}
\newcommand{\gsim}{\lower.5ex\hbox{\gtsima}}
\newcommand{\ltsima}{$\; \buildrel < \over \sim \;$}
\newcommand{\gtsima}{$\; \buildrel > \over \sim \;$}
\def \etal      {\hbox{et al.} }
\newcommand{\subfind}{\texttt{\footnotesize SUBFIND }}
\newcommand{\CMBFAST}{\texttt{\footnotesize CMBFAST }}

\shortauthors{BUSHA, EVRARD, \& ADAMS}
\shorttitle{The Power Spectrum and Structure\\} 
\begin{document} 

\title{The Asymptotic Form of Cosmic Structure: \\
Small Scale Power and Accretion History}

\author{Michael T. Busha, August E. Evrard\altaffilmark{1},
Fred C. Adams\altaffilmark{1}}

\affil{Michigan Center for Theoretical Physics \\
Physics Department, University of Michigan, Ann Arbor, MI 48109\\
mbusha, evrard, fca@umich.edu} 

\altaffiltext{1}{Astronomy Department, University of Michigan, Ann Arbor, MI 48109} 

\begin{abstract} 

We explore the effects of small scale structure on the formation and
equilibrium of dark matter halos in a universe dominated by vacuum
energy.  We present the results of a suite of four N-body
simulations, two with a \LCDM initial power spectrum and two with
WDM-like spectra that suppress the early formation of small
structures.  All simulations are run into to far future when the
universe is $64 \hinv$Gyr old, long enough for halos to essentially
reach dynamical equilibrium.  We quantify the importance
of hierarchical merging on the halo mass accretion history, the
substructure population, and the equilibrium density profile.  We
modify the mass accretion history function of \citet{W02} by
introducing a parameter, $\gamma$, that controls the rate of mass
accretion, $d\ln M / d\ln a \propto a^{-\gamma}$, and find that this
form characterizes both hierarchical and monolithic formation.
Subhalo decay rates are exponential in time with a much shorter time
scale for WDM halos.  At the end of the simulations, we find truncated
Hernquist density profiles for halos in both the CDM and WDM
cosmologies.  There is a systematic shift to lower concentration for
WDM halos, but both cosmologies lie on the same locus relating
concentration and formation epoch.  Because the form of the density
profile remains unchanged, our results indicate that the equilibrium
halo density profile is set independently of the halo formation process.  

\end{abstract}

 
\section{INTRODUCTION} 

Our understanding of cosmology has consolidated significantly in
recent years and we are now approaching percent-level estimates of the
most important cosmological parameters (e.g., \citealp{WMAP3}).  Using
these parameters, numerical simulations of large scale structure
provide excellent descriptions of the distribution and properties
of dark matter halos.  The next open question is to understand why
non-linear structure takes the form that it does --- a form that has been
predicted by simulations and confirmed, more or less, by observational
data.  In particular, we would like to know how and why dark
matter halos attain a nearly universal form for their density profiles,
as first described by Navarro, Frenk, \& White 1997 (hereafter
NFW). One aspect of this issue is understanding the importance of the
method of mass accretion:  How much does the final structure depend on
accreting mass as virialized clumps as opposed to a continuum of
diffuse material and how effectively does violent relaxation
\citep{LyndenBell} erase the memory of this accretion process.

The basic process for the buildup of structure in our cold dark matter
(CDM) dominated universe is the hierarchical merging of collapsed
structures (see, e.g., \citealp{PS74, Aarseth79, Blumenthal84,
Davis85}).  This process creates small halos early in the universe
which merge with each other while accreting material from their
surroundings, eventually creating the large cluster-sized structures
of today through a ``bottom up'' process. 

For a time it was thought that light neutrinos might dominate the mass
density, forcing galaxy formation to occur through a ``top
down'' process (e.g., \citealp{Bond83}).  In such a hot dark matter
model, perturbations on small scales are washed out by free streaming,
preventing the formation of the early low-mass seeds of hierarchical
structure formation.  Dark matter halos still form, but the first
objects are large cluster-mass halos \citep{Zeldovich70,
Doroshkevich75}.  While popular in the 1970's because simulations reproduced
the outline of the cosmic web that surveys were just beginning to map
out \citep{Thompson78}, hot dark matter models have been ruled out
based on observations of the galaxy distribution \citep{White83}.
Tuning the free-streaming mass-scale leads to WDM possibilities which
suppress density perturbations below some (typically dwarf galaxy
sized) scale.  The most immediate effect
of this suppression is to reduce the number of small halos and
subhalos existing in large halos. Several numerical studies of WDM
cosmologies have been carried out (e.g., \citealp{Evrard92, Avila01, Bode01,
Cloin00, White00, Knebe02, Knebe03}), mostly in an attempt to explain
the apparent lack of substructure in our local group as compared to
predictions from \LCDM simulations. 

Although CDM and hierarchical merging has emerged as the standard
paradigm for structure formation, and is the key ingredient in setting
the distribution of dark matter  halos, it is still uncertain how
important this process of mass accretion is in setting the internal
properties of dark matter halos.  While some studies indicate that
the accretion of substructures plays a significant role in setting the
inner slope of the radial halo density profile \citep{Ma04},
simulations of monolithic collapse events in WDM scenarios produce
halos with global properties unchanged relative to similar structures
in CDM cosmologies \citep{Evrard92, Moore99}.  By truncating
the initial power spectrum in an otherwise CDM simulation at some
scale $k_c$ (with corresponding mass scale $M_c =
4/3\pi^4k_c^{-3}\bar{\rho}$), one can mimic WDM cosmologies and test
the importance of hierarchical growth for establishing halo
structure.  Compared to the CDM model, where mass is continually
accreted in dense clumps, WDM cosmologies
have an early period of monolithic collapse, where large halos
form out of a smooth background and relax with many fewer disruptions
due to merger events.  If the halo is in a dense enough region, this initial
collapse is followed by hierarchical accretion.  Previous simulations
of this process \citep{Moore99, Bode01, Avila01} have shown that the
resulting density profile is virtually unchanged for halos well above
the truncation scale.

Halos in a WDM cosmology are effectively a
re-scaled versions of the first halos expected in a CDM cosmology.
Most physical CDM candidates (i.e., SUSY-LSP's) have some
intrinsic velocity dispersion, washing out perturbations on very small
scales (much smaller than we are able to simulate on cosmological
scales), effectively truncating the power spectrum at some very large
$k$ (see \citealp{Diemand05, Gao05} for discussions of such
simulations).  In this manner, studying WDM cosmologies can provide clues
to the earliest collapse of CDM structures. 

The discovery of the accelerated expansion of the universe (Perlmutter
\etal 1998), potentially caused by a non-zero cosmological constant
(the \LCDM model), provides us with a mechanism for studying the true
equilibrium structure of cosmic halos.  Previous work
\citep{Adams97, Nagamine03, Busha03, Busha05} has shown that, in the relatively
near cosmic future of a \LCDM universe, structure formation and halo
growth will come to a rapid end.  This happens around $a \sim 3$, when
$\ov \approx 1$.  At this time, exponential deSitter expansion causes
mergers and accretion to stop and pushes existing halos further and further
away from each other.  From $a \sim 3$ beyond, no matter is
left for halos to accrete, and they relax toward their asymptotic
equilibrium state in effective isolation.  These late-time halos
experience no disruption from mergers or other events that frustrate
the equilibrium, in contrast to the situation at the present epoch.

To gain insight into the question of the origin of the internal
structure of dark matter halos, we have performed a set of simulations
of cosmic structure, using both a standard \LCDM and truncated WDM-like
power spectra.  Simulations are run into the far future, allowing
halos to relax toward equilibrium configurations. 
Our numerical simulations are described in \S 2.  In \S 3, we compare
the differences in the halo distribution for the two cosmologies,
concentrating on the mass function, evolution of the power spectrum,
and the formation of WDM halos with mass well below the truncation
scale.  In \S 4 we compare the internal structure of the halos in the two
cosmologies, including the mass accretion histories (MAHs), the
distribution of substructure, and the halo density profiles. Results
and their implications are summarized and discussed in \S 5.

\section{SIMULATIONS} 

We simulate the formation of dark matter halos in $\Lambda$-dominated
CDM and WDM-like cosmologies with a suite of dark matter N-body
simulations using the publicly available TreePM code Gadget 2.0
\citep{Gadget2}. We use two \LCDM and WDM simulations with
different mass resolutions for a total of four large-scale
cosmological simulations.  The simulation pairs at each mass
resolution were created with the same initial phases so that the
the large-scale environment would be unchanged.  All simulations model
a patch of space in a flat universe with current matter density $\om =
0.3$, vacuum density $\ov = 0.7$, Hubble parameter $H_0 = 70 {\rm
km~ s}^{-1}\mpc^{-1}$ and power spectrum normalization
$\sigate = 0.9$, values consistent with the first year release of WMAP
measurements of the CMB power spectrum \citep{WMAP}.  The lower
resolution simulations model periodic cubes of side length $L = 200
\hinv \mpc$ containing $N_p = 256^3$ particles of mass $3.97 \times
10^{10} \hinv \msol$ and gravitational softening length $\epsilon_p = 40
\hinv \kpc$.  The higher resolution simulations use cubes of side
length $L = 50 \hinv \mpc$, containing $256^3$ particles with mass
$6.20 \times 10^8 \hinv \msol$ and softening length $\epsilon_p = 10
\hinv \kpc$.
The softening scales quoted here correspond to their values at the
present epoch and are held constant in comoving space for $a
< 1$, but become physical lengths for $a > 1$ to prevent structures
from being over-softened due to the exponential increase in $a$ during
the deSitter phase.  Table \ref{table:Simulations} lists
these simulation parameters.  All simulations were
started at redshift $z = 19$ (scale factor $a = 0.05$) and were run
into the far future, $a = 100$.  Although the starting redshift is
somewhat late, it is consistent with the analysis by \citet{Power} for
a simulation of our resolution, and should present no problem for the
late-time results we are primarily concerned with.  We store a total
of 300 outputs equally spaced in $\log(a)$ for each simulation.  At $a
= 100$ the universe is about $64\hinv$Gyr old and structure formation
has preceded to completion in a \LCDM cosmology \citep{Busha03,
Nagamine03}.   The simulations were run on 16 nodes of a dual-Opteron
Beowulf cluster at the University of Michigan.

\begin{table}
\caption{Simulation parameters:}
\begin{center}
\begin{tabular}{ccccc}
\tableline
\tableline
{ $L [\hinv\mpc]$} & {$M_p [\hinv\msol]$} & {$\epsilon_p [\hinv\mpc]$}
& {$N_p$} \\
\tableline
200 & $3.97 \times 10^{10}$ & 0.04 & $256^3$\\
50 & $6.20 \times 10^{8}$ & 0.01 & $256^3$\\
\tableline
\label{table:Simulations}
\end{tabular}
\end{center}
\end{table}

\begin{figure}
\includegraphics[width=3.5in,keepaspectratio=true]{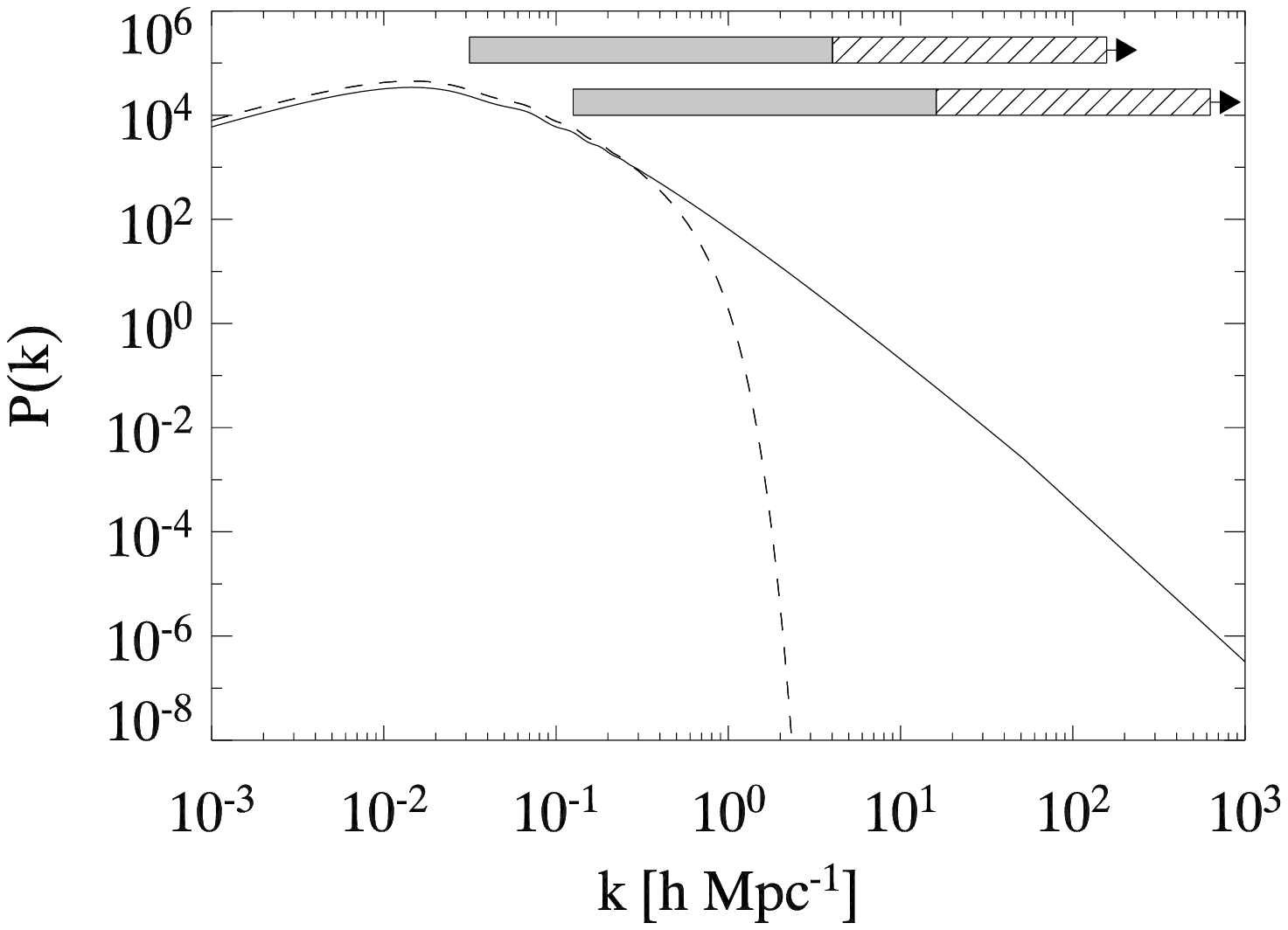}
\figcaption
{The input power spectra for the \LCDM (sold line) and WDM (dashed
line) simulations.  The \LCDM spectrum is calculated using \CMBFAST.
The WDM model adds an exponential cutoff to the \LCDM model at a mass
scale of $8.09 \times 10^{13} \hinv\msol$.  Both spectra were
normalized so that $\sigate = 0.9$. The light gray boxes represent the
range between the fundamental and Nyquist frequencies for the small
(top) and large (bottom) volume runs.  The hatched-line boxes show the
range between the Nyquist and softening frequencies for $a \le 1$.
For $a > 1$, the softening is constant in the physical frame and its
corresponding wavenumber grows with $k \propto a$ in this comoving
representation. 
\label{fig:PowerSpectrum}
}
\end{figure}

\begin{figure*}
\plotone{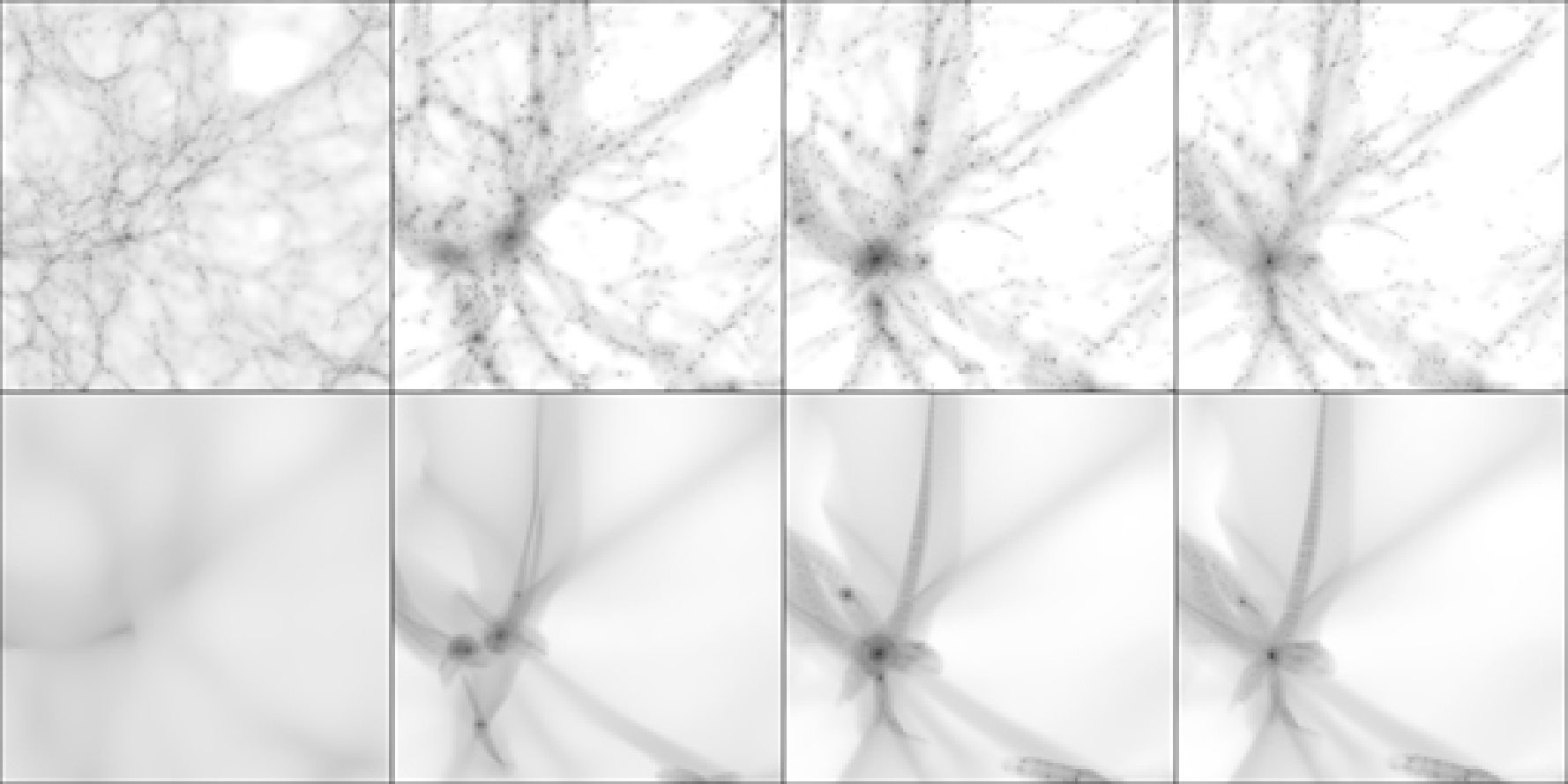}
\figcaption
{The density fields of comoving slices of the cosmic web at $a=0.3,
1, 3$, and 100 (columns, left to right) of a \LCDM (top) and WDM
(bottom) cosmologies from our small volume simulations.  The side
length for each image is $35\hinv\mpc$, with thickness
$5\hinv\mpc$. The grey-scale is proportional to $\log(\rho /
\bar{\rho})$. 
\label{fig:WebComoving}
}
\end{figure*}

The initial power spectrum for our CDM simulations was set using
\CMBFAST \citep{CMBFAST} in accordance with WMAP year 1 data
\citep{WMAP}.  Our second, WDM-like, simulation used a truncated
version of this initial power spectrum,
\be
P_{t}(k) = CP_{0}(k)e^{-(k/k_c)^2},
\label{eqn:PSpec}
\ee
where $k_c$ is the truncation scale and $C$ a normalization
coefficient that allows us to set $\sigate$.  Figure \ref{fig:PowerSpectrum}
plots our input spectra, with $P_0$ as the solid line and $P_t$ as
the dashed line. For our simulations, we choose a truncation scale of
$k_c = 0.511 h\rm{Mpc^{-1}}$, which corresponds to a mass scale of
$M_c = 8.09\times 10^{13} \hinv \msol$ (2037 particles for our lower
resolution simulations and 130,396 at the higher resolution).  This
truncated spectrum was re-normalized to $\sigate = 0.9$ so that high
mass halo abundances would be similar.  As noted earlier, phase
information was retained for each resolution pair, resulting in
similar large scale structures (Figure \ref{fig:WebComoving}) and
allowing us to identify corresponding halos in the two cosmologies. 

Equation (\ref{eqn:PSpec}) represents a transfer function that
differs from the standard transfer function for a WDM cosmology.  
Generally, the mass of the WDM particle, $m_{W}$ sets a free
streaming length, $R_f = 0.2(\Omega_{W}h^2)^{1/3}(m_{W}/1{\rm
keV})^{-4/3} \mpc$ \citep{Sommer01}, which approximates the WDM power
spectrum through the relation \citep{Bardeen86}
\be
P_{WDM}(k) = \exp \left[ -kR_f - (kR_f)^2 \right] P_{CDM}(k).
\label{eqn:PWDM}
\ee
This spectrum has a slightly more gradual cutoff than
equation (\ref{eqn:PSpec}).  It should be noted that we refer to our
truncated models as ``WDM'' cosmologies even through they were not
created using this transfer function.  For reference, our truncated
model most closely approximates a WDM cosmology with $m_{WDM} = 0.13
\rm{keV}$ and $R_f = 1.6 \mpc$.

\begin{figure*}
\plotone{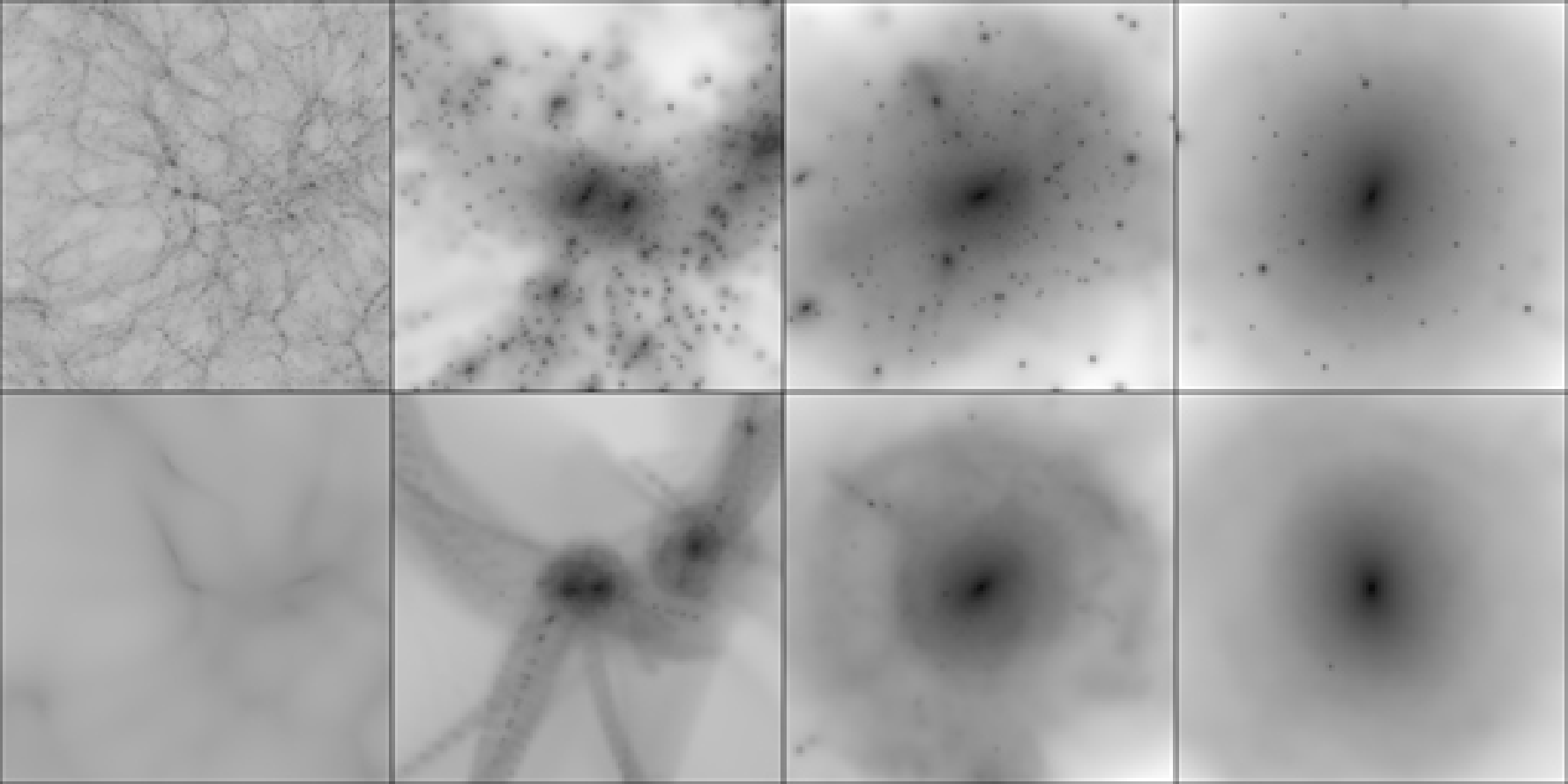}
\figcaption
{The density field around a large halo  at
$a=0.3, 1, 3$, and 100 (columns, left to right) from our small volume
simulations.  The top row shows the largest halo from the \LCDM
cosmology ($\mvir = 5.38\times10^{14}\hinv\msol$ at $a = 100$) and
the bottom row shows the corresponding WDM halo ($\mvir = 5.24 \times
10^{14} \hinv\msol$
at $a = 100$).  The
side length for each image is $12\hinv\mpc$, with thickness
$6\hinv\mpc$ in physical units.  The grey-scale is proportional to
$\log(\rho / \rho_c)$.  
\label{fig:WebPhysical}
}
\end{figure*}

The evolution of the resulting density fields of the simulations are
shown in Figures \ref{fig:WebComoving} and \ref{fig:WebPhysical}.
Figure \ref{fig:WebComoving} shows a slice the comoving density field
from our smaller volume CDM and WDM simulations at $a = 0.3, 1, 3$,
and 100.  The differences between these two models are striking,
especially at $a = 0.3$, where the CDM cosmology exhibits a
well-formed web with an abundance of small halos.  The WDM cosmology, 
in contrast, has a mostly uniform density, with only one
visible halo and a handful of weak filamentary structures making up
the cosmic web.  A clear cosmic web does rapidly develop in the WDM
simulation, however, and by $a = 1$ similar large scale structures are
present in both cosmologies, even though there is a
strong suppression of small halos in the WDM filaments.  By $a =
3$, the large scale density field is set and undergoes little evolution
from $a = 3$ to 100.  Once the cosmological constant
becomes dominant the growth function saturates, ending halo formation
and freezing the comoving web.  Halos continue to contract in this
comoving picture, and by $a = 100$ they consist of small, tightly
bound knots along and at the intersection of filaments. 

Figure \ref{fig:WebPhysical} shows the evolution of the density
field in a fixed physical region.  Here, we focus on the evolution
of a particular CDM halo with mass $\mvir = 5.38 \times 10^{14} \hinv
\msol$ and its counterpart in the WDM cosmology (see \S 3.3).  While
much of the late time growth is identical (such as the
major merger around $a = 1$), the initial formation processes differ
substantially.  At $a = 0.3$, there are many low mass CDM progenitor
halos present, while the WDM halo looks like a weak (uncollapsed)
perturbation in an otherwise smooth background.  This difference
is manifest throughout all plotted epochs by the persistent lack of
substructure in the WDM halo at the three later epochs. 

\begin{figure*}
\plotone{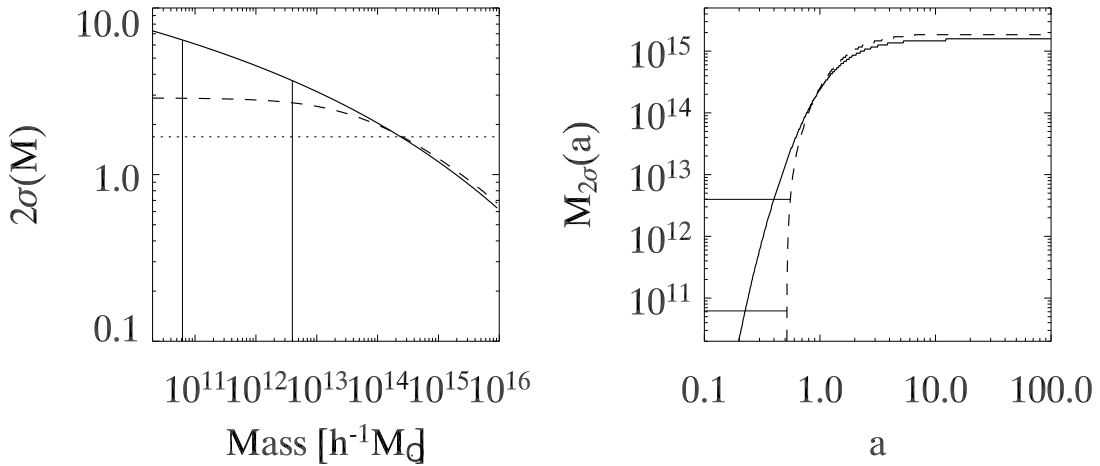}
\figcaption
{{\it Left:} Rare ($2\sigma$) perturbation amplitude as a function of
mass at $a = 1$ for the \LCDM (solid line) and WDM (dashed line)
models.  The vertical lines show the mass scales for 100 particles
at the two resolutions. {\it Right:} Characteristic collapsed mass as
a function of scale factor for the \LCDM (solid line) and WDM (dashed line)
cosmologies. The solid horizontal lines show the mass scales for 100
particles at the two resolutions.  
\label{fig:MSigStar}
}
\end{figure*}

One way to quantify the expected suppression of hierarchical buildup
is to look at the critical mass scale, $\mstar$, where one expects a
perturbation to go non-linear and collapse, defined through the relation
\be
\label{eqn:MStar}
\sigma[\mstar] = {\delta_c \over D(a)}.
\ee
Here, $\sigma(M) = (2\pi)^{-3} \int_0^{\infty}
P(k)\widetilde{W}^2(M,k) d^3k$, $\widetilde{W}$ is the Fourier
transformation of the top-hat window function, $D(a)$ is the linear
growth function, and $\delta_c = 1.686$ is the linearly extrapolated
criterion for collapse of an overdense perturbation \citep{PS74,
Peebles80}.  The factor$D(a)$ can be calculated numerically from the
expression 
\be
D(a) \propto H(a)\int_{0}^{a} {da' \over [a'H(a')]^3},
\ee
and is normalized such that $D(1) = 1$.  The left panel of Figure
\ref{fig:MSigStar} shows the amplitude of $2\sigma(M)$ perturbations
as a function of $M$ at the present epoch.  The horizontal dotted line
shows the critical scale, $\delta_c = 1.686$.   The right panel of
Figure \ref{fig:MSigStar} shows the evolution of $M_{2\sigma}(a)$ ---
the mass of a $2\sigma$ perturbation that goes nonlinear as a function
of cosmic time. In both cases, the solid line represents the full
power spectrum model and the dashed line represents the truncated
power spectrum model.  We consider $2\sigma$ perturbations because, in our
WDM cosmology, $1\sigma$ perturbations do not collapse until
$a = 1.33$ and are prevented from growing substantially by the
cosmological constant.  The normalization $\sigate = 0.9$ boosts the
power at large mass ($M > 10^{14} \hinv \msol$) in the WDM model
relative to the CDM case and causes the largest perturbations to
collapse slightly earlier and with larger asymptotic masses in the WDM
model.  We also expect no structures in the WDM cosmology for $a <
0.4$, agreeing with Figure \ref{fig:WebPhysical} in which the most
massive halo of the simulation has yet to form a coherent structure
prior to this epoch.  

Dark matter halos in our simulations are identified using a standard
friends-of-friends (FOF) algorithm with linking length 0.15 times the
inter-particle spacing.  Halo centers are identified as the most
bound particle of the resulting group.  As a mass measure capable of
spanning both early and late times we use $\mvir$, the mass of all
particles inside a sphere of radius $\rvir$ with over-density
$200$ times the critical density, $\rho_c$.  In previous work, we
found the $\mvir$ provides a good proxy for the asymptotic halo mass
and is roughly half the value of the ultimate halo mass, $\mvir \simeq
0.5M_{halo}$ \citep{Busha05}.  The halo velocity is defined to be the
center of mass velocity of all particles within $\rvir$.

Subhalos were identified using the \subfind routine
\citep{Subfind}.  This routine works on top of an
FOF group and identifies density maxima within halos from an SPH
smoothing kernel that uses the distance to the $32^{nd}$-nearest
neighbor to obtain local density estimates.  Subhalos are then
selected as locally overdense regions containing at least 20 bound
particles.  The largest subhalo identified by \subfind is actually the
host halo of the FOF group.   At $a = 100$, this host halo should
correspond to the actual equilibrated halo as defined in
\citet{Busha05}, minus any locally bound subhalos.  A comparison
between these mass estimates shows good agreement, with the masses
agreeing to within 1\%.  

Throughout this paper, we use several scale radii.  Halo sizes are
defined using $\rvir$ (see above) and $r_{halo}$, the spherical radius
containing all bound particles (which is only defined for $a \gsim 5$,
see \citealt{Busha05}).  Additionally, we fit NFW and Hernquist density
profiles to our halos, 
\bee
\rho_{NFW} = {4 \rho_s \over r/r_s(1+r/r_s)^2},\label{eqn:nfw}\\
\rho_{Hern} = {\rho_0 \over r/r_c(1+r/r_c)^3},
\eee
which adds the scales $r_s$ (the radius where
the best fit NFW profile has logarithmic slope $-2$) and $r_c$ (where
the best fit Hernquist profile has slope $-2.5$).  Generally, we find
that $r_{halo} = 4.6\rvir$ and $r_s = 0.4r_c$ (see \S4.3, \citealp{Busha05}).  

\section{COMPARISONS OF THE DARK MATTER DISTRIBUTION}

In this section we compare properties of the distribution of dark
matter halos in our simulations, including the evolution of the
power spectrum, the halo mass function, the correspondence between CDM
and WDM halos, and the formation of WDM halos below the
truncation scale.  

\subsection{Evolution of the Power Spectrum}

\begin{figure*}
\plotone{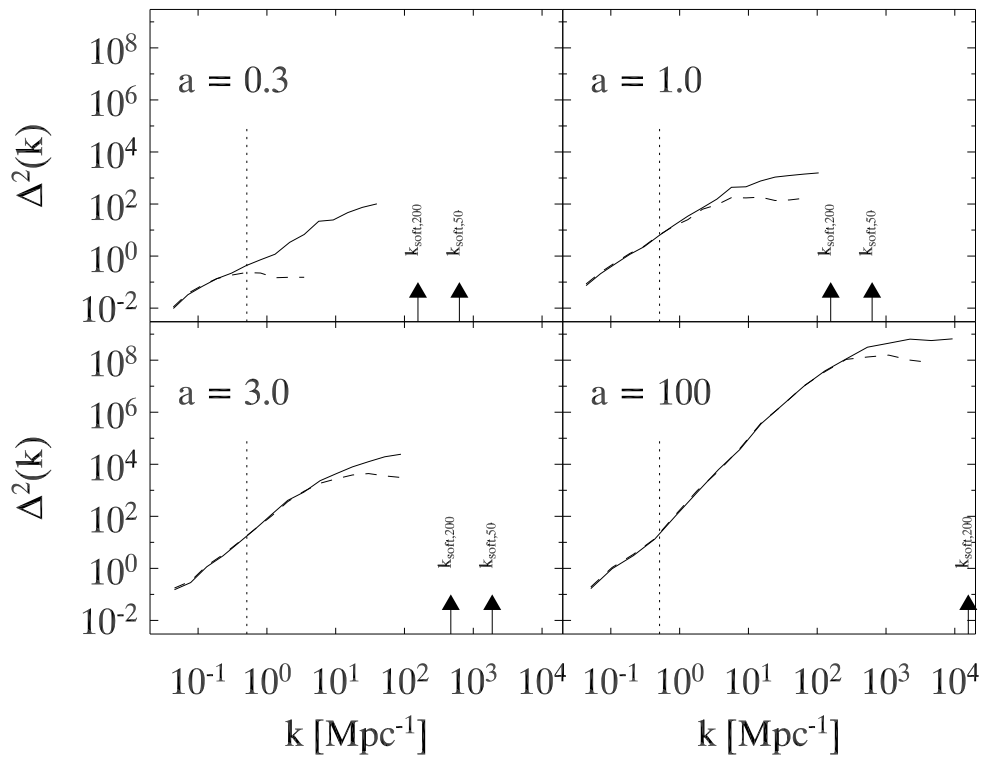}
\figcaption
{The evolution of the power spectrum for halos in CDM (solid lines) and
WDM (dashed lines) cosmologies.  The spectra are shown
at $a = 0.3, 1, 3,$ and 100.  Power spectra from the small
and large volume realizations are laid on top of each other.  The
dotted vertical lines represent the truncation scale, $k_c$, and the
arrows show the softening scales for the large and small volume runs.
The spectra are plotted from the fundamental frequency of the
simulation volume to the frequency where the shot noise becomes
comparable to the measured power, generally $\sim 0.1k_{soft}$. 
\label{fig:PSpecEvln}
}
\end{figure*}

Figure \ref{fig:PSpecEvln} shows the evolution of the power spectrum
for all four of our simulations at the epochs $a = 0.3, 1, 3$, and
100.  The solid lines represent the CDM cosmologies and the
dashed lines the WDM spectrum.  The power spectrum is shown in
dimensionless units, $\Delta^2(k) \propto k^3P(k)$.  Power spectra for
the large and small volume simulations are combined, allowing us to
probe a larger range in $k$.  The arrows represent the softening lengths
for the large and small volume simulations at the plotted epoch.  The
collapse of non-linear structure creates substantial power beyond the
Nyquist frequency, which we measure using the tiling method of
\citet{Jenkins98}.  The spectra are plotted from the fundamental
simulation frequency out to a wavenumber where the shot noise of a
Poisson distribution of particles becomes comparable to the measured
power.

At $a = 0.3$, when non-linear structure formation is in
its early stages, the power spectrum of the WDM model is heavily
truncated above $k_c$.  By the present epoch, much of this
suppression has disappeared due to power transfer from collapsing
structures, and the WDM cosmology matches the CDM model almost
perfectly at low and intermediate wavenumber, up to an order of
magnitude above $k_c$. Relatively little happens to the power spectrum
beyond $a = 1$.  As noted earlier, the dominance of $\Lambda$ halts
the growth of structure beyond $a \sim 3$ and causes the power
spectrum to freeze after only a modest amount of additional evolution.
Since Figure \ref{fig:PSpecEvln} plots the power spectrum in comoving
space,  beyond $a = 1$ the expansion of the universe transfers power
to larger scales with no real change in the shape of the spectrum.
For most of the measurable range, $0.1 < k < 100 \rm{h Mpc}^{-1}$, the
power spectrum is well characterized by the expected power law,
$\Delta^2 \propto k^3$, caused by the freezing of structure in an
expanding universe.  Collapse, however, has not managed to elevate the
WDM power spectrum up to that of the CDM model at all scales by $a =
3$, and a small suppression of power at the highest wavenumbers persists at
$a = 100$.  This suppression is due to the lack of low mass halos.  If
we measure the halo power the spectrum using only halos with $\mvir >
M_c$ we recover spectra that are identical at the few percent level,
which indicates that the distribution of halos on scales greater than
$M_c$ is statistically unchanged.  

The evolution of the power spectrum in WDM cosmologies was studied
previously by \citet{White00} and \citet{Knebe02}.  These studies
present results similar to ours.  By $a \sim 0.5$, non-linear collapse
has boosted the truncated portion of the power spectrum substantially.
agreeing with a \LCDM model for $k < 10 h\mpc^{-1}$ with only a
slight suppression for larger $k$.  

\subsection{Mass Function}

\begin{figure*}
\plotone{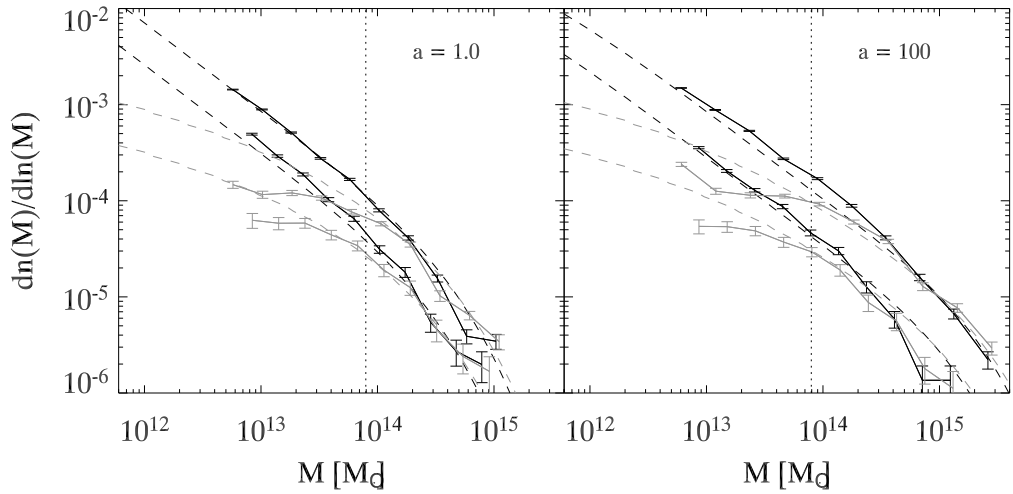}
\figcaption
{The mass function for the CDM and WDM cosmologies (dark and light
curves) at $a = 1$ and $100$ for our larger volume simulations.  The
upper curves show the FOF(0.164) mass function and the lower curves
the SO(200) mass function.  The FOF(0.164) function has been offset
vertically for clarity.  The vertical dotted line shows the truncation scale.
Error bars assume Poisson statistics.  
\label{fig:MassFunction}
}
\end{figure*}

Figure \ref{fig:MassFunction} shows the mass function for our large volume
CDM and WDM simulations at $a = 1$ and 100.  Errors are calculated
assuming Poisson statistics and the vertical dotted line is the
truncation scale, $M_c$.  
Here, the dark lines represent the \LCDM model and the light lines
the WDM version.  For comparison, the figure also shows the Jenkins
Mass Functions (JMF, \citet{Jenkins01}) as dashed lines.  The JMF is
defined via
\bee
{dn_{JMF}(M,a) \over d\ln(M)} & = & A {\bar{\rho} \over M} {{\rm d} \ln
\sigma^{-1}(M,a) \over d \ln M} \times \nonumber\\
& & \exp[-|\ln \sigma^{-1}(M,a) +
B|^{\epsilon}], 
\label{eqn:Jenkins}
\eee
where $A, B, {\rm and ~} \epsilon$ are fitting
parameters.  Two mass values are used in this plot to compare with
published JMF parameters: FOF masses with a linking length $b = 0.164$
(upper curves, \citealt{Jenkins01}) and $\mvir$ from a spherical
overdensity groupfinder (SO, lower curves, \citealt{Evrard02}).  The fitting
parameters are listed in Table \ref{table:JMF}.  For the $a = 100$
spherical overdensity JMF, we used the $\Omega_M = 0$ parameters from
\citet{Evrard02}.  Although not shown, The Sheth \& Tormen mass
function with published parameters \citep{ST} agrees quite well with
the JMF FOF mass function at all epochs.  While the agreement between
FOF masses in the CDM
cosmology and equation (\ref{eqn:Jenkins}) is good for
$a = 1$, a substantial mass excess is present in the mass range
$10^{13} < M_{\rm FOF} < 5 \times 10^{14} \hinv\msol$ at $a = 100$.
Additional simulations of this cosmology confirm that this excess is
significant and not a result of one particular realization.  The use
for FOF masses in measuring the mass function does create difficulties
in the far future because the FOF groupfinder identifies isosurfaces
relative to the background matter density, which is dropping
rapidly from the exponential expansion.  Compared with the critical
density, the physical density isosurface identified by a FOF
groupfinder is $\rho_{\rm FOF} = [(a^6\Omega_{m,0})/(\Omega_{m,0} +
a^3\Omega_{\Lambda,0})] b^{-3}\rho_{crit}$.  By $a = 3$, $\rho_{\rm
FOF} = 3.5 \rho_{crit}$, which includes a substantial amount of
material outside the virialized region of a halo that is unbound and
being pulled away by the Hubble flow \citep{Busha05}.  At late
epochs, $M_{\rm FOF}$ is a poor proxy for the actual (virialized) mass
of a halo.  The $\mvir$ mass function, however, does not suffer from
this defect.  The $a = 1$ result agrees with \citet{Evrard02} to $\sim
20\%$ in number, approximately
the quoted statistical accuracy.  The parameters are slightly off,
however, due to insufficient mass resolution of the simulation.  The
$a = 100$ mass function does provide a substantially better agreement
than the FOF mass function, fitting to within about $2\sigma$ at all
masses.  

\begin{table}
\caption{Mass Function Parameters:}
\begin{center}
\begin{tabular}{ccccc}
\tableline
\tableline
{Mass Function} & {$A$} & {$B$} & {$\epsilon$} \\
\tableline
JMF -- FOF(0.164) & 0.301 & 0.64 & 3.88\\
JMF -- SO(200, $a=1$) & 0.220 & 0.73 & 3.86\\
JMF -- SO(200, $a=100$) & 0.199 & 0.76 & 3.90\\
\tableline
\label{table:JMF}
\end{tabular}
\end{center}
\end{table}

The mass function for the WDM cosmology exhibits a striking
suppression for low masses, beginning slightly above the the
truncation scale, $M_c$ (dotted vertical line).  An unexpected upturn
appears in the WDM model at the lowest masses ($\lsim 50$ particles)
for FOF halos.  \citet{Bode01} and \citet{Knebe02} claim that this
behavior is a result of physical halos forming through fragmentation
from Jeans instability.  However, such an upturn is not
present in the SO mass function and we show evidence in the appendix
that it is actually a numerical artifact of the FOF groupfinder.  We
have also calculated JMF and ST fits for our WDM cosmology, which do
not fit as well as in the \LCDM cosmology.  Both the JMF and ST
mass functions strongly over-predict the abundance of halos with
masses $M < M_c$.  The poor fit in the range $10^{13} \hinv\msol < M <
8\times10^{13}\hinv\msol$ should not be surprising because equation
(\ref{eqn:Jenkins}) was motivated by a perturbation collapse threshold,
similar to the Press-Schechter (1974) model, which uses spherical
collapse to determine a collapse epoch.  As we discuss in the next
section, WDM halos with mass below $M_c$ form out of larger mass
perturbations that do not follow this model, at odds with
the assumptions of Press-Schechter.  

\subsection{Halo Correspondence}

\begin{figure}
\includegraphics[width=3.5in,height=2.7in]{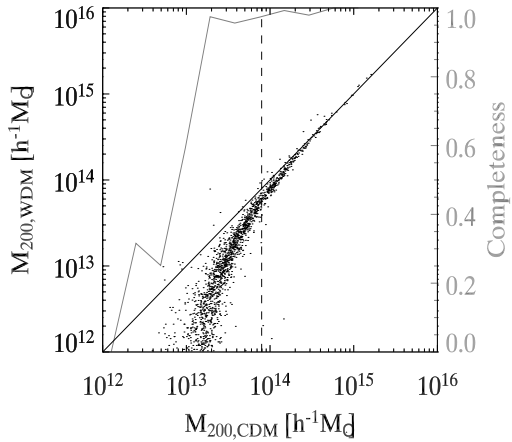}
\figcaption
{A comparison of the $\mvir$ values for corresponding halos in the
\LCDM and WDM runs at $a = 100$.  The dashed vertical line denotes the
truncation scale.  Halos with $\mvir \gg \mtrunc$ have roughly the
same mass with a scatter of about 6\%.  The light line shows the
probability for finding a WDM halo corresponding to a \LCDM halo of a
given mass. 
\label{fig:CrossHalos} 
}
\end{figure}

Because we used the same phases in constructing the initial
conditions, we can cross-match halos in the CDM
and WDM cosmologies using a Lagrangian scheme.  We select a FOF halo
in one cosmology (usually
\LCDM) and identify for the largest halo in the other cosmology
containing at least 50\% of particles of the selected halo.  When
starting with a \LCDM halo, we don't allow any two WDM halos to be
identified with the same \LCDM halo. This simple method is robust for
massive systems, and corresponding halos are found for 98\% of all
halos with $M_{200} > M_c = 8.08\times10^{13}\hinv\msol$.  The
``missing'' halos are lost because the smoothing of the power spectrum
causes distinct halos in the CDM cosmology to form as single
halos in the WDM run.  Our requirement that each WDM halo have only a
single CDM counterpart prevents all but the most massive of these CDM
halos from having a WDM counterpart.  If we relax this correspondence
requirement, all CDM halos with $M_{FOF} > M_c$ have WDM
counterparts.  Figure \ref{fig:CrossHalos} shows the masses of
corresponding halos in the WDM and \LCDM models.  At higher masses
($\mvir > 2\mtrunc$) the masses are similar, $M_{200,CDM} \approx
M_{200,WDM}$ with about a 6\% scatter.  As the mass falls below the
truncation scale, $M_c$, the WDM halos become less and less massive
relative to their CDM counterparts, eventually disappearing
altogether.  There are a few extreme outliers from the general
relationship.  Halos with a low $M_{WDM}/M_{CDM}$ (the three halos
with $M_{200,CDM} \sim 10^{14}\hinv\msol$ and $M_{200,WDM} \sim
2\times 10^{12} \hinv\msol$) are CDM halos in filaments that never
completely collapse in the WDM cosmology.  Most of their particles are
located in a
spray along the filament, but there is a small WDM halo in the filament
with $\sim 90\%$ of its members in the CDM halo.  The halo with the
$M_{200,WDM} \gg M_{200,CDM}$ is a rare occurrence where many small
CDM halos ($\sim 10$) had merged into a single, much more massive halo
in the WDM cosmology.  

Also plotted in Figure \ref{fig:CrossHalos} is the completeness
function for identifying corresponding halos.  Virtually all \LCDM
halos with $\mvir > \mtrunc$ have corresponding WDM halos, but the
completeness function drops very rapidly for $\mvir < M_c$, the mass
range where halo formation is strongly suppressed.  

\subsection{Sub-Truncation Scale Halos In WDM}

One surprising observation from the WDM simulations is that many halos
form with mass scale
$\mvir$ well below the truncation scale, $M_c$.  Although physical halos
with $\mvir < M_c = 8.08\times10^{13}\hinv\msol$ account for only a
few percent of all collapsed mass, these halos represent a substantial
population by number.   \citet{Knebe03} have shown that these halos
appear in well-defined filaments and argue that they are the result of Jeans
instability \citep{BT}.  Our results confirm that these halos form
in filamentary structures, and we find that many of these halos have
CDM counterparts, indicating that these low mass WDM halos are
not numerical artifacts.  

In order to understand how these sub-truncation scale halos
form, we identify particles in small halos ($\mvir < M_c$) in
our small volume simulations at $a = 100$ and trace them back to
the initial conditions.  We then locate the density peak nearest the
center of mass of this distribution and measure the mass of the
perturbation by finding the radius where spherical overdensity drops
below $\delta_c D(a)$.  For all halos in the WDM simulation, the
perturbation mass is in the range $(0.2 - 2) \times 10^{14} \hinv
\msol = 0.25 - 2.5 M_c$.  In particular, no halos form out of
perturbations substantially smaller than our truncation scale, but
sub-truncation scale halos do form out of large perturbations
that do not fully collapse.  The CDM simulation, by comparison, has
small halos forming out of perturbations with masses
anywhere between the final halo mass, $M_{halo}$, and $2\times
10^{14}\hinv\msol$, with the bulk of the halos (80\%) forming from
perturbations with mass less than 4 times $M_{halo}(100)$.  The
distribution of perturbations with initial masses greater than $M_c$,
however, is almost identical to that for the WDM cosmology.  Low
mass WDM halos are simply large perturbations that do not collapse
completely, and behave the same as in a CDM cosmology.  Such
a trend was noted by \citet{Katz93}, and our WDM halos seem to be an
extreme case of the tendency for halos to form objects with a
substantially different mass than their initial spherical collapse
prediction. 

\section{COMPARISON OF HALO PROPERTIES}

This section compares properties of individual halos in our CDM and
WDM cosmologies.  In particular, we focus on the mass accretion
histories, subhalo abundance, and density profiles.  In spite of some
substantial and fundamental differences in the first two properties,
the form of the radial density profiles is unchanged between the
CDM and WDM cosmologies.  Furthermore, halo concentrations follow
the same relation with formation epoch in both models. 

\subsection{Mass Accretion Histories}

One property for which we expect a clear difference between
CDM and WDM halos is the halo mass accretion history (MAH).
For the WDM model, the reduced merger activity should result in
a smoother MAH since mass is primarily accreted in the form of diffuse
material.  The suppression of power at large $k$ also alters the
characteristic collapse mass at low mass/early times (Figure
\ref{fig:MSigStar}, right panel).  The implication is that
halos will form later and more rapidly (in the sense of a larger
$d\ln M / d \ln a$) in a WDM cosmology.

The MAHs are measured using a halo's most massive
progenitor, where a progenitor is any halo at a preceding output in which
at least 50\% of the FOF particles end up in the subsequent halo.
When comparing halos between the two runs, we first select CDM halos
from a given mass range and then select either their WDM counterparts
(as in Figure \ref{fig:CrossHalos}, ignoring CDM halos that have no
match) or WDM halos from the same mass ranges.  These selection
methods are nearly degenerate for $\mvir > M_c$, as shown in Figure
\ref{fig:CrossHalos}. At lower CDM halo masses $M_{200,{\rm WDM}}$ is
strongly suppressed, so the selection methods differ substantially.
Generally, we prefer to consider halos of similar final masses so that
we do not have to worry about halos with no counterpart.  

\begin{figure}
\includegraphics[width=3.5in,height=3in]{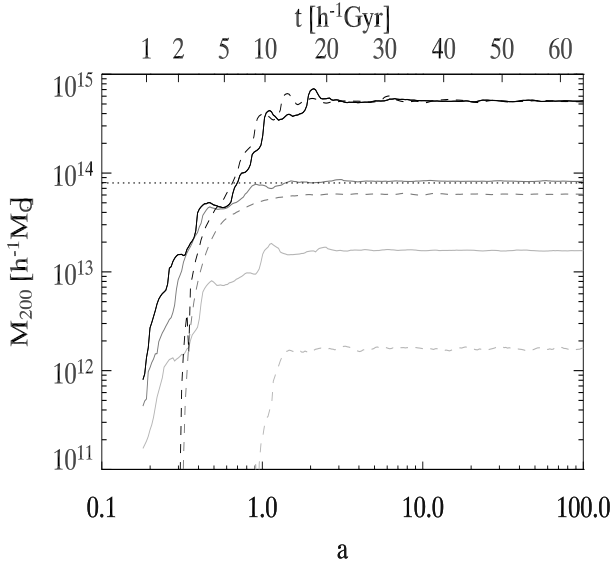}
\figcaption
{The accretion history for three individual halos in
the full run (solid curves) and their corresponding halos in the truncated run
(dotted curves).  The halos were selected with $\mvir \gg M_c$ (dark
curves), $\mvir \sim M_c$ (medium curves), and $\mvir \ll M_c$ (light curves)
in the run with the full power spectrum.  The dotted horizontal line
represents $M_c$.  
\label{fig:MAcc1}
}
\end{figure}

Figure \ref{fig:MAcc1} shows MAHs for three halos in the
full run (solid curves) and their matched halos in the truncated
run (dashed curves).  These individual halos were selected from our
smaller volume run to have $\mvir \gg M_c$ (dark curves), $\mvir \sim
M_c$ (medium curves), and $\mvir < M_c$ (light curves) at $a = 100$ in the
CDM cosmology.  As expected, the halos in the truncated model have
smoother MAHs during the initial halo growth phase and form slightly
later, with these effects becoming more pronounced for smaller halos.
The overall shape of the MAHs for the halos with $\mvir \gg M_c$ is
remarkably similar.  While the most massive halo in the truncated
model has no progenitor before $a \sim 0.2$, it grows quickly and
catches up with the CDM halo by $a = 0.7$.  Afterwards, the two halos
evolve almost identically, even undergoing the same major mergers
around $a = 1$ and 2.  These mergers happen slightly earlier in the
WDM cosmology due to the increase in power at this scale from our
normalization $\sigate = 0.9$, which effectively starts the ``cosmic
clock'' for these large halos at a later time.  In
contrast with the these late-time similarities, the MAHs at $a < 0.7$
are substantially different.  In the WDM scenario, a large mass
perturbation will collapse more or less as a unit, as soon as it goes
non-linear. This collapse creates a phase of smooth, rapid mass growth
with $d\ln M_{{\rm WDM}} / d\ln a \gg d\ln M_{{\rm CDM}} / d\ln
a$. Once the halo reaches $M_{200,{\rm WDM}} \approx M_c$, the halo
is the approximate size of its counterpart in the CDM cosmology and
$d\ln M_{200} / d\ln a$ drops to match the rate of the CDM halo.  The
WDM halo then begins to accrete mass as already-collapsed clumps in a
quasi-hierarchical fashion.  The CDM halos with masses $\mvir \le
M_c$, in contrast, have corresponding WDM halos that never accrete
mass in virialized clumps. The MAHs of such WDM halos are much
smoother and accrete the bulk of their mass in a single period of rapid
accretion. 

\begin{figure}
\includegraphics[width=3.5in,height=3.8in]{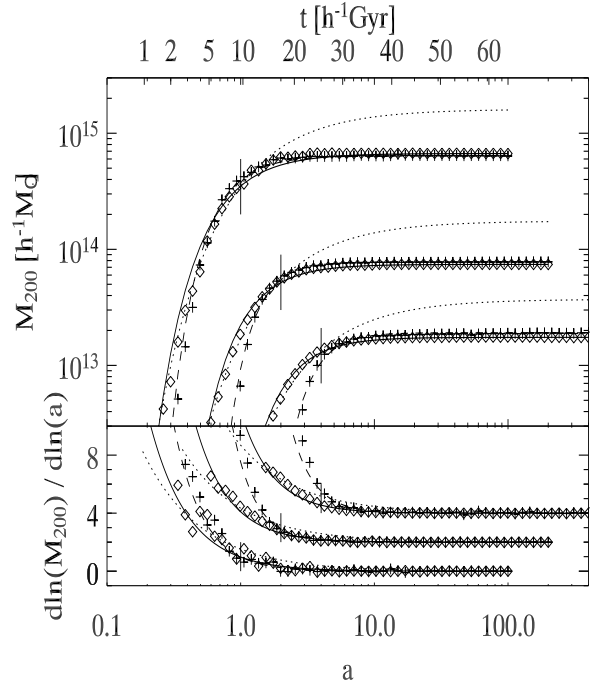}
\figcaption
{{\it Top Panel: } Average MAHs for CDM (diamonds) and  WDM halos
(plus symbols) from our large volume simulations.  The halos are selected
from the mass ranges $\mvir = (1 - 4) \times 10^{13}, (0.5 - 1.3)
\times 10^{14},$ and $ > 4 \times 10^{14} \hinv\mpc$.  The curves are
offset in $a$ by a factor of 2 (intermediate mass range) and 4 (low
mass range) to make them easier to distinguish, but the short vertical
lines represent $a = 1$ for all mass ranges.  The dotted curves are
fits to equations (\ref{eqn:w02}) for the CDM halos and the solid and
dashed curves are fits to equation (\ref{eqn:Mofa}) for the CDM and
WDM cosmologies. {\it Bottom Panel: } The mass
growth rates, $d\ln(M_{200})/d\ln(a)$, of the halos and plotted above
and their fits to equation (\ref{eqn:Mofa}).  Curves are offset in
both the horizontal and vertical directions to make them more
distinguishable.  
\label{fig:MAcc2} 
}
\end{figure}

Figure \ref{fig:MAcc2} shows ensemble average MAHs for halos from our
large volume simulations.  Here, the diamonds represent CDM halos
and the crosses are WDM halos.  Halos from different mass ranges have been
offset in time to make the figure more readable.  Both the CDM and WDM
halos are selected to lie in the mass ranges $\mvir = (2 - 4) \times
10^{13} \hinv\msol, (0.5 - 1.3) \times 10^{14} \hinv\msol, {\rm and }
> 4 \times 10^{14} \hinv\msol$ at $a = 100$. Figure \ref{fig:MAcc2}
shows many of the same trends observed in Figure \ref{fig:MAcc1}. For
the most massive halos, the two cosmologies again show mass
equality around $M_{200,{\rm CDM}} \approx M_{200,{\rm WDM}} \approx
M_c$, with the WDM halos accreting mass significantly faster before
this time.  For $M_{200} \le M_c$, WDM halos form later and more
rapidly than CDM halos with similar mass.  Also shown in the bottom
panel of this figure are the growth rates, $d\ln(M_{200})/d\ln(a)$, of
the halos and their fits to equation (\ref{eqn:Mofa}). 

Wechsler \etal (2002, hereafter W02) proposed a fitting formula for
the MAH of a halo up to the present epoch of the form 
\be
M(a) = M_0e^{-(a_{c,W02}/a_0)S(a_0/a - 1)}.
\label{eqn:w02}
\ee
The free parameter $S$ in this equation is used only in defining
$a_{c,W02}$, the creation epoch for the halo, when $d \ln M / d
\ln a = S$.  We choose to follow their convention and adopt $S = 2$.  This
formula is fit to our CDM MAHs over the range
$a = 0.2 - 1.0$ (but continued out to $a = 100$) and plotted as the
dotted curves in Figure \ref{fig:MAcc2}.  In general, the fit is good
for both CDM and WDM (not shown) models for $a < 1$, but overestimates
halo masses by a factor of 2 in the CDM run and more than an order of
magnitude for the WDM run at late times.  If the fit is calculated for
the full range, $0.2 < a < 100$, the late time asymptote is correct,
but $d\ln M_{200} / d\ln a$ is substantially lower than observed for
either cosmology at all epochs.  This behavior is probably an
indirect result of the coincidence problem --- the surprising
observation that we live during the relatively short epoch where
$\Omega_M \approx \Omega_{\Lambda}$. The fit works well for $a
< 1$, even when calculated for just a fraction of the
region and then extrapolated. Although equation (\ref{eqn:w02}) was
created for halos in a \LCDM universe, $a = 1$ is not much later than
the equality epoch, $a_{eq} = 0.75$, when $\om = \ov$.  Once $\Lambda$
becomes the dominant component of the universe, the growth function
quickly saturates and halos cease to grow \citep{Busha05}.
Consequently, equation (\ref{eqn:w02}) approaches its asymptote much
more  slowly than halos feeling the full effects of a dominant
cosmological constant.  To capture the full histories in both
cosmologies, we propose a generalization of the form 
\be
M(a) = M_0e^{-(a_c/a_0)^\gamma{S \over \gamma}((a_0/a)^\gamma - 1)},
\label{eqn:Mofa}
\ee
where $\gamma$ is the rate index which sets the mass growth rate
through the relation $d\ln M / d\ln a = S(a_c/a)^{\gamma}$.  This
variable is introduced such that $a_c$ is still the epoch where $M(a)$
has a logarithmic slope of $S$, but $\gamma$ sets how quickly a halo
grows and asymptotes to its equilibrium mass.  When $\gamma = 1$ this
generalized form reduces to equation (\ref{eqn:w02}).  We
expect to recover $\gamma > 1$, which corresponds to more rapid
formation and faster asymptote behavior (see Figure \ref{fig:MAcc2})

\begin{figure}
\includegraphics[width=3.5in,height=2.in]{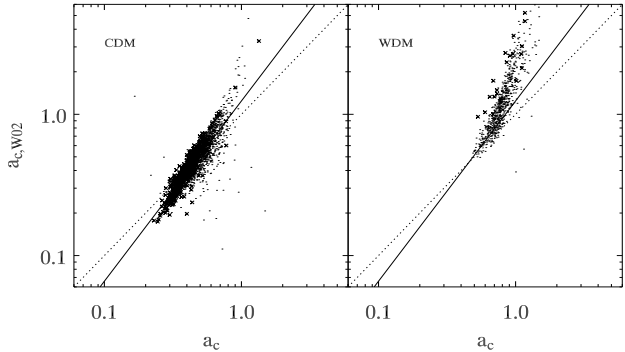}
\figcaption
{Comparison of the formation epochs defined by W02 (equation
[\ref{eqn:w02}]) and our generalization (equation [\ref{eqn:Mofa}]).
The left panel represent halos from the CDM run and the right panel
shows WDM halos.  Dots are halos from the large volume run and crosses are
from the small volume run.  The dotted line shows an exact
correspondence and the solid black line is a polynomial fit to the
CDM halos.
\label{fig:acComp}
}
\end{figure}

The fits to equation (\ref{eqn:Mofa}) in Figure \ref{fig:MAcc2} (solid
and dashed curves for CDM and WDM halos) closely follow the measured
MAHs. They provide good agreement over all epochs and work
equally well for both CDM and WDM halos.  Although not
shown, the errors for this fit are generally $\lsim 5-10\%$, depending on
the number of halos we average over. Figure \ref{fig:acComp} compares
$a_{c,W02}$, from equation (\ref{eqn:w02}), and $a_c$ from our modified
form, equation (\ref{eqn:Mofa}), with dots representing halos from the large
volume simulations, and crosses are from the small volume simulations.  In
the CDM cosmology (left panel), $a_{c,W02} = 2.7a_c^{1.4}$ with a 27\%
scatter. The few WDM halos with $a_{c,W02} < 1$ are also well
described by this relation.  In both cases, there are several halos
with $a_{c,W02} > 1$. These are typically halos that first appear
around $a \ge 0.7$ and have rapid growth phases, powered either by
major mergers or the collapse of a sub-truncation scale perturbation
in the WDM cosmology. These formation epochs, based on equation
(\ref{eqn:w02}), appear unphysical since they give
formation epochs in the period of exponential expansion when halo
growth has stopped.  Equation (\ref{eqn:Mofa}) resolves this issue by
increasing the halo rate index, resulting in a substantially lowered
$a_c$ which pulls these halos significantly above the measured linear
relation between $a_c$ and $a_{c,W02}$. 

\begin{figure*}
\plotone{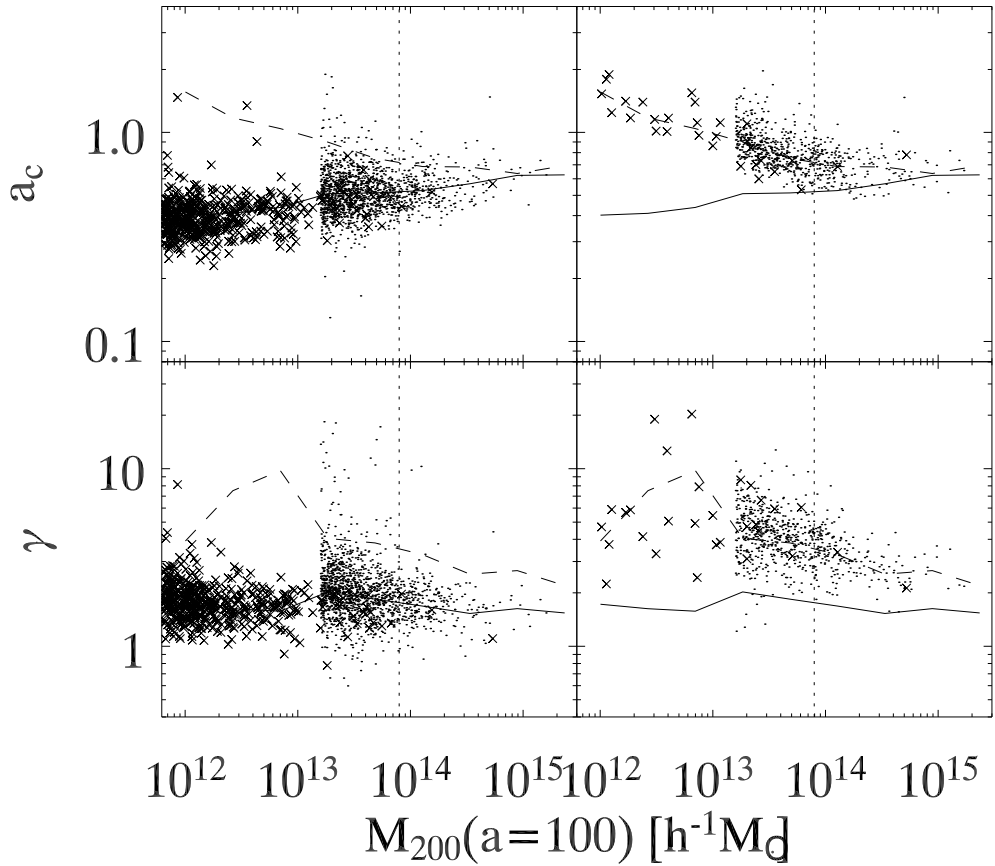}
\figcaption
{{\it Top Panels:} The dependence of halo's formation epoch ($a_c$
in equation [\ref{eqn:Mofa}]) on mass.  The left panel shows halos from the CDM
run, and the right halos from the WDM cosmology.  Dots represent
halos from the large box run, and crosses are from the small box run.  The
solid and dashed curves show the trend of $a_c$ mass for the CDM and
WDM cosmologies, and the vertical dotted line is $M_c$. {\it Bottom
Panel:} Same as the top panel, but now plotting the behavior of
$\gamma$, the amplification factor of equation
(\ref{eqn:Mofa}), as a function of mass. 
\label{fig:acofM}
}
\end{figure*}

Figure \ref{fig:acofM} plots the variation of $a_c$ (top panels) and
$\gamma$ (bottom panels) with mass for CDM (left panels) and WDM
(right panels) halos.  Again, dots represent halos from the large
volume realizations, and crosses are halos from the smaller volumes. 
Common to both the left and right panels are average trend lines for
the CDM (solid curves) and WDM (dashed curves) halos.  The plot shows
all halos from our simulations that are well resolved at the end of
the simulation ($\mvir > 400$ particles for the large volume
realizations and $\mvir > 1000$ particles for the smaller volume
realizations --- see appendix for a further discussion).  At the high mass end,
$\mvir > 2M_c =  2 \times 10^{14} \hinv \msol$, the average $a_c$'s
differ by only $\sim 10\%$, less than the scatter for either
cosmology.  The rate index, $\gamma$, however, is about 50\% higher in
the WDM cosmologies at these high masses, reflecting the steeper
MAHs presented in right side of Figure \ref{fig:MAcc2}. 

As mass decreases, both $a_c$ and $\gamma$ behave differently in the
two cosmologies.  
The formation epoch decreases with mass in the CDM runs
(in accordance with ``bottom up'' structure formation) but actually
increases with mass in the WDM cosmologies.
The number of WDM halos that exist with $M_{200} \ll M_c$ --- all with
later formation times than halos with mass greater than $M_c$ ---
again suggests that most small halos form through an
instability of regions inside larger structures.  This claim is
consistent with our picture of sub-truncation scale halos forming
through incomplete collapse of larger perturbations. For CDM
halos, we again see the presence of the hierarchical structure
formation from the fact that $\gamma$ is relatively constant
throughout the entire mass range, $\gamma \approx 2$ with modest
scatter.  In contrast, the WDM cosmology has $\gamma$ increasing with
lower masses, roughly as $\gamma \propto M_{200}^{-0.2}$.  The spray
of particles with $a_c$ and $\gamma$ much larger than the averages in
the CDM cosmologies is a result of the difficulty in measuring these
parameters for poorly resolved halos.  The MAH cannot be
measured accurately for halos that do not grow substantially above our
resolution limit, and consequently our fit parameters have large
uncertainties.  The mean relation is actually within these
uncertainties for all the low mass halos.  

\begin{figure}
\includegraphics[width=3.5in,height=3.5in]{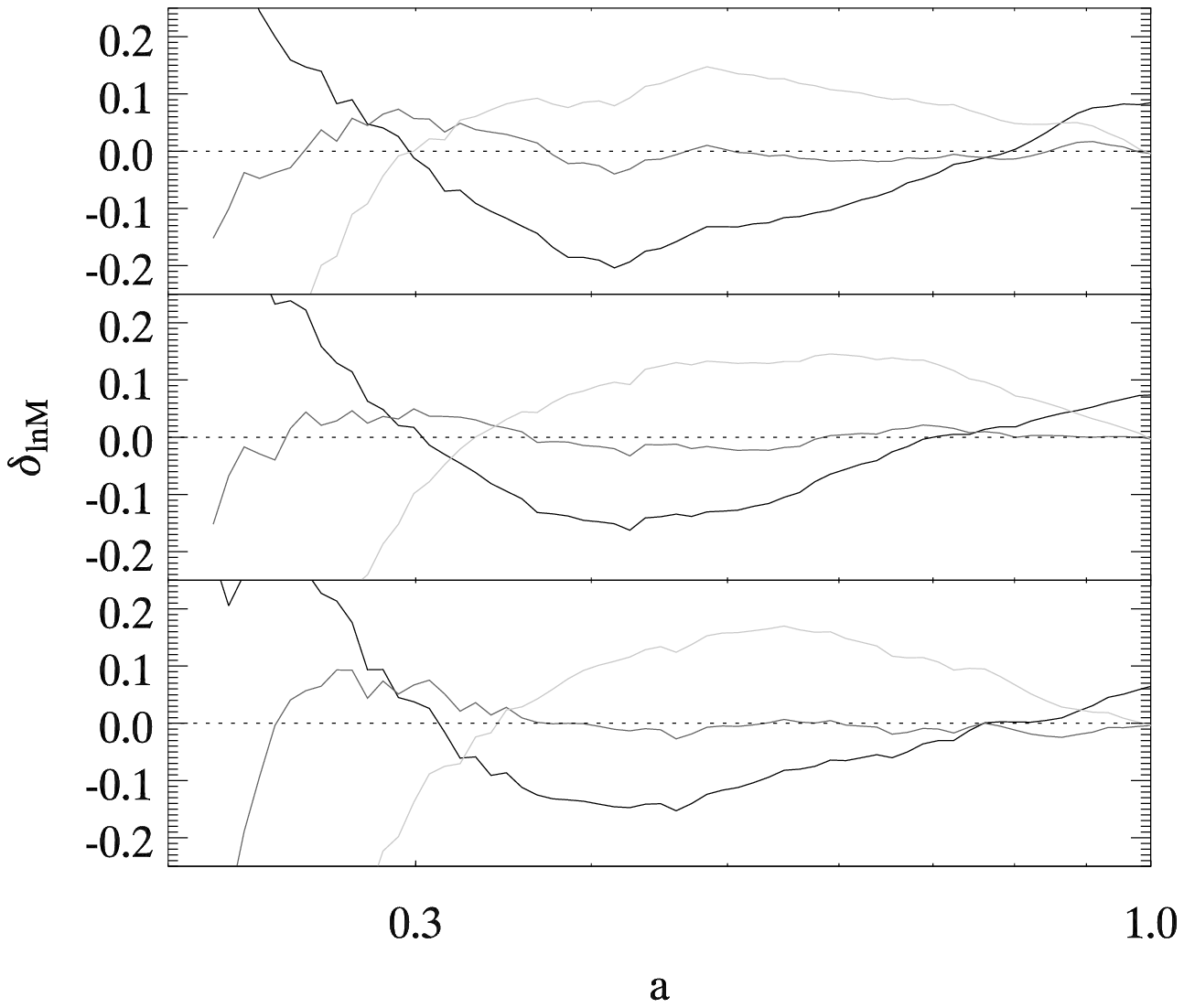}
\figcaption 
{Residuals of average MAHs to various fits.  The dark lines are the
residuals to equation (\ref{eqn:Mofa}) fit in the range $a = 0.3 -
100$.  The medium lines also use equation (\ref{eqn:Mofa}), but the
fit is calculated over $a = 0.3 - 1$.  The light lines fit the model
of \citet{W02}, equation (\ref{eqn:w02}), to the range $a = 0.3 -
1$.  The halos are in the mass ranges (top to bottom) $\mvir = (0.6 -
1.2) \times 10^{14}, (3 - 6) \times 10^{13}, (1.3 - 3) \times 10^{13}
\hinv \msol$.
\label{fig:CompareMofA}
}
\end{figure}

We also compare our modified MAH fitting formula with the model of
W02 in the range of halo formation up to the present epoch in
Figure \ref{fig:CompareMofA}.  In this figure, we compare the
residuals of fits of the average MAH to equation (\ref{eqn:Mofa}), fit
from $a = 0.3 - 100$ (dark line) and $a = 0.3 - 1$ (medium line) with
equation (\ref{eqn:w02}) fit from $a = 0.3 - 1$ (light line) for
different mass ranges taken from our CDM simulations.  The RMS
values for the residuals in $\ln M$ are shown in Table
\ref{table:residuals}.  While the rms residual between our modified
fitting formula from $a = 0.3 - 100$ is approximately a factor of 2
lower than the W02 model for all mass ranges, equation
(\ref{eqn:Mofa}) fit from $a = 0.3 - 1$ offers a substantial
improvement over equation (\ref{eqn:w02}), decreasing the rms residual
by a factor of 5 or more at the plotted masses.  
The introduction of $\gamma$ is apparently an important correction
for the early time MAH growth as well as the late time asymptote.
Table \ref{table:acFits} lists the values for $a_c$ and $\gamma$ for
the profiles of Figure \ref{fig:CompareMofA}.  The fits are robust in
the sense that $a_c$ does not change substantially depending on how
the MAH is calculated. In all cases, however, the best fit returns
$\gamma$ substantially larger than 1, and may have some mass
dependence.  The exact value of $\gamma$ depends strongly on the fit
range, with fits out to $a = 100$ requiring a larger value in order
for the MAH to asymptote properly. These larger values of $\gamma$ in
turn push $a_c$ slightly earlier. Fitting in the range $0.3 \le a \le
1$, results in $\gamma \approx 1.6$ with values for $a_c$ that differ
from those of the W02 model by only 4\%.  While our average
halo MAHs all appear to have $\gamma > 1$, there is a much larger
spread when considering fits for individual halos. Here, $\gamma$
ranges from $0.2 - 10$ and the fits have residuals that are typically
20\% lower than those of equation (\ref{eqn:w02}).

\begin{table}
\caption{RMS residuals of MAH fits:}
\begin{center}
\begin{tabular}{cccc}
\tableline
\tableline
{ $\mvir [\hinv\msol]$} & {Equation (\ref{eqn:Mofa}):} &
{Equation (\ref{eqn:Mofa}):} & {Equation (\ref{eqn:w02}):}
\\
{ } & {$a = 0.3 - 100$} & {$a = 0.3 - 1$} & {$a = 0.3 - 1$}
\\
\tableline
$(0.6 - 1.2) \times 10^{14}$ & 0.14  & 0.044 & 0.21\\
$(3 - 6) \times 10^{13}$ & 0.14 & 0.036 & 0.31\\
$(1.5 - 3) \times 10^{13}$ & 0.15 & 0.088 & 0.43\\
\tableline
\label{table:residuals}
\end{tabular}
\end{center}
\end{table}

\begin{table*}
\caption{MAH Parameters:}
\begin{center}
\begin{tabular}{cccccc}
\tableline
\tableline
{ $\mvir [\hinv\msol]$} & {$a_{c,W02}$} & {$a_c$} & {$a_c$} & {$\gamma$} &
{$\gamma$}
\\
{ } & {$0.3 - 100$} & {$0.3 - 1$} & {$0.3 - 1$} & {$0.3 -
100$} & {$0.3 - 1$}
\\
\tableline
$(0.6 - 1.2) \times 10^{14}$ & 0.54 & 0.47 & 0.52 & 2.0 & 1.5\\
$(3 - 6) \times 10^{13}$ & 0.53 & 0.50 & 0.52 & 2.1 & 1.6\\
$(1.5 - 3) \times 10^{13}$ & 0.48 & 0.48 & 0.50 & 2.3 & 1.8\\
\tableline
\label{table:acFits}
\end{tabular}
\end{center}
\end{table*}

\subsection{Halo Substructure}

In this section, we compare the subhalo distribution of our CDM
and WDM cosmologies, considering only halos from our smaller volume
simulations.  While these simulations do not contain a statistically
large number of halos, the larger volume simulations do not have the
necessary resolution to accurately describe the subhalo population.  

Not surprisingly, the most dramatic difference between the subhalo
populations of our CDM and WDM halos is their abundance.
In the CDM simulation, the average number of subhalos with $M > 1.24
\times 10^{10} \hinv \msol$ (20 particles) is roughly proportional to
the mass of the host halo, $\bar{n}_{subs} \propto M_{200}$ at $a = 1$.
For host halos of all masses, approximately 10\% of the host mass is in
bound substructures at this epoch, a value consistent with previous
studies \citep{Klypin99}.  In contrast, for the WDM cosmology
$\bar{n}_{subs} \propto M_{200}^{0.4}$, with only about 5\% of the host mass
in bound subhalos.  By $a = 100$ in the CDM cosmology, the slope of
the number of subhalos with mass has not changed substantially,
$\bar{n}_{subs} \propto M_{200}^{1.2}$, but many subhalos have been
destroyed, and only 0.3\% of the mass of an average halo is contained
in substructure.  The steepening of the slope is caused by a more
efficient destruction of subhalos in hosts with lower masses.  Small
halos today contain smaller, more weakly bound, subhalos than larger
hosts.  Additionally, these small halos undergo fewer future mergers
to replenish their subhalo population.  By contrast, with the
exception of the single largest WDM halo, none of the WDM halos
contain any substructures at $a = 100$. Even this largest halo
has only $\bar{n}_{subs} = 2$, as opposed to $\bar{n}_{subs} =
137$ in its CDM counterpart.  Surprisingly, the shape of the subhalo
mass function, $dn_{sub}(M_{sub}) / d\ln(M_{sub}/M_{Host}),$ does not change
substantially between today and $a = 100$.   The mass function is
truncated at the high mass end ($M_{sub}/M_{Host} > 0.01$) and is
slightly steepened.  This is due to the increased effect of dynamical
friction on more massive objects, pulling them towards the center of
the halo where they are more easily disrupted and stripped of
mass.  

\begin{figure}
\includegraphics[width=3.5in,height=3in]{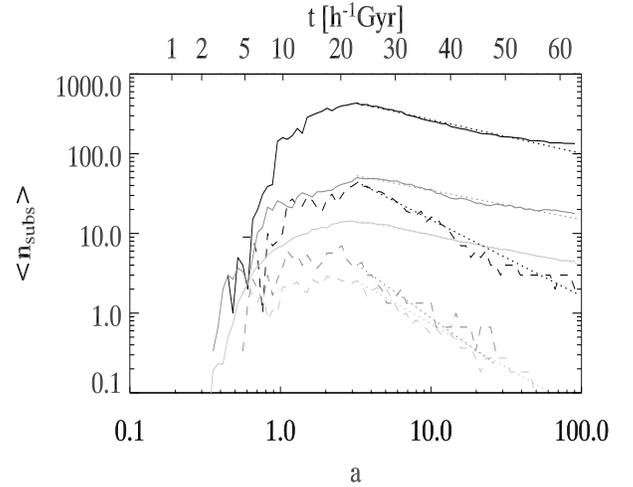}
\figcaption
{The evolution of the average number of subhalos in host halos of
various masses.  The solid curves indicate CDM halos, and the dashed
WDM halos.  At $a = 100$, the halos have masses $\mvir =
(2 - 4)\times10^{13} {\rm (light ~line),~ } (0.5 - 1.3) \times10^{14}
{\rm (medium ~line), and~ } > 4\times10^{14} \hinv\msol {\rm (dark
~line)}$.  The dotted lines show fits to equation (\ref{eqn:nsuba}).
\label{fig:nSubsA}
}
\end{figure}

Figure \ref{fig:nSubsA} shows the evolution of the average
number of subhalos in halos of various masses.  CDM and WDM halos
of all sizes show similar evolution in the average number of
subhalos.  For $a \le 3$, mergers create substructure, resulting in an
increasing $\bar{n}_{subs}(a)$.  After mergers end, $a \approx 3$, no
new subhalos are accreted and
existing subhalos gradually fall inward and are disrupted due to
dynamical friction and tidal forces.  The number of subhalos in the
WDM cosmology, however, drops much more rapidly than in the CDM
cosmology.  The primary reason for this difference is the lower
binding energy of the subhalos caused by the later formation epoch of
low-mass WDM halos.  When these halos accrete onto more massive halos,
they are much more prone to disruption and consequently have a shorter
life.  While WDM subhalos are formed with a lower binding
energy/density, most other properties, including their average
velocity, velocity dispersion, and mass, differ very little between
the CDM and WDM models.  The only other systematic difference is the
average distance a subhalo lives from the center of its host --- WDM
subhalos tend to live further out.  This is also related to the lower
binding energy of WDM subhalos, since they are more easily disrupted
when they move closer to center of their host.  As a measure of
subhalo destruction, we have fit the evolution of $\bar{n}_{sub}$
during the late-time deSitter expansion (when the host halos are no
longer being disrupted by mergers) to an exponential decay,
\be
\bar{n}_{sub}(a) = \bar{n}_m e^{-\alpha' (t-t_m)} = \bar{n}_m
(a/a_m)^{-\alpha}, 
\label{eqn:nsuba}
\ee
where the subscript $m$ denotes the epoch where the subhalo population
is at its maximum, and $\alpha'$ is the subhalo decay rate, with
$\alpha' = H\alpha = \alpha 0.0856{\rm Gyr}^{-1}{\rm h}$ for our
$\Lambda$-dominated cosmology ($H = H_{\infty}$).  For our CDM
halos, we get $\alpha = 0.38 \pm 0.03$, while for the WDM halos there
is a much more rapid decay, $\alpha = 1.1 \pm 0.2$, yielding subhalo
half-lives of $21 \pm 2$ and $7.4 \pm 1 \hinv$Gyr, respectively.
Figure \ref{fig:nSubsA}
makes it appear that equation (\ref{eqn:nsuba}) fits better for lower
mass halos.  This, however, is an artifact of the fact that we have
only a few high mass halos in each simulation, giving us poor
statistics.  

\subsection{Density Profile}

\begin{figure}
\includegraphics[width=3.5in,height=3.7in]{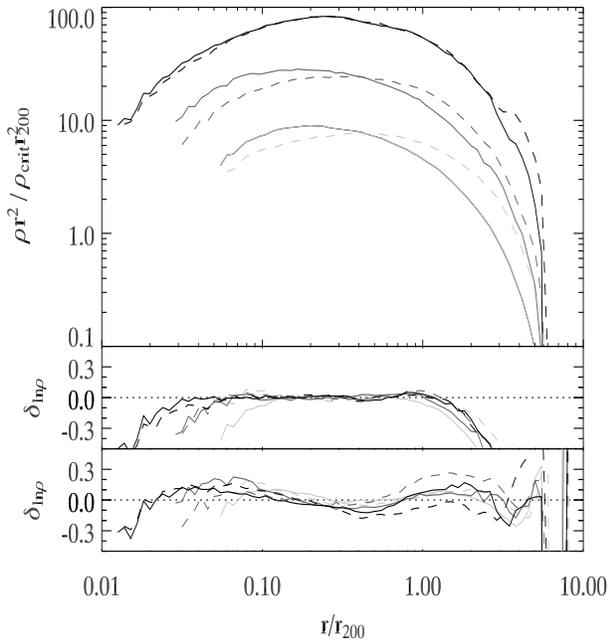}
\figcaption
{{\it Top Panel:} Average density profiles times $r^2$ for all
bound material in halos at $a = 100$ from our smaller volume runs.
Halos selected according to the mass ranges $\mvir = (2 - 4) \times
10^{13} {\rm (light ~line),~ } (0.6 - 1.3) \times 10^{14} {\rm (medium
~line), and ~ } > 4\times10^{14} \hinv\msol {\rm (dark ~line)}$, with
the solid and dashed lines representing CDM and WDM halos.  The
profiles have been offset from each other to make them easier to
see. {\it Middle Panel:} Residuals of NFW fits to the above six
profiles. {\it Bottom Panel:} Residuals of truncated Hernquist fits
(equation [\ref{eqn:hern}]) to the above six profiles.
\label{fig:Rho}
}
\end{figure}

The radial density profile is one of the most fundamental halo properties.
Previous WDM studies \citep{Moore99, Cloin00, Bode01} indicate that WDM
halo density profiles do not differ substantially from their CDM
counterparts, and our simulations support this finding.  Figure
\ref{fig:Rho} shows average density profiles for all particles
bound to halos from the CDM and WDM cosmologies at $a = 100$ from
our smaller volume runs. The solid lines are CDM halos from the mass
ranges of Figure \ref{fig:MAcc2}, and the dashed lines are WDM halos.
The different mass ranges have been offset and the density multiplied
by $r^2$ to make the differences between the various profiles more
visible.  The halos can be described by an NFW profile,
equation(\ref{eqn:nfw}) for the range $0.05\rvir \lsim r \lsim \rvir$.
Here, $r_s$ and $\rho_s$ are the the NFW scale radius and density.
The middle panel of Figure \ref{fig:Rho} shows the residuals to the
NFW fits for the nine plotted profiles, which generally fall in the
$\sim 5\%$ range.  

Previous work \citep{Busha05} has shown that well before $a = 100$,
halos in a $\Lambda$-dominated universe develop clear edges at $r
\approx 4.6\rvir$.  Beyond $\rvir$, the NFW profile is much
too shallow and the density profile for all bound material is better
fit by a truncated Hernquist profile, 
\be
\rho = {\rho_0 \over (r/r_c)(1+r/r_c)^3}
e^{(-r/r_{halo})^{5.6}},
\label{eqn:hern}
\ee
where $\rho_0$ and $r_c$ are the central density and core radius, and
$r_{halo}$ is the radial extent of all bound halo material.  Residuals
to this fit are shown in the bottom panel of Figure \ref{fig:Rho}.
This fit has larger residuals than NFW for $r < r_{200}$, but has much
smaller residuals at larger radii and provides a good description of
the halo out to its actual edge.  

\begin{figure}
\includegraphics[width=3.5in,height=2in]{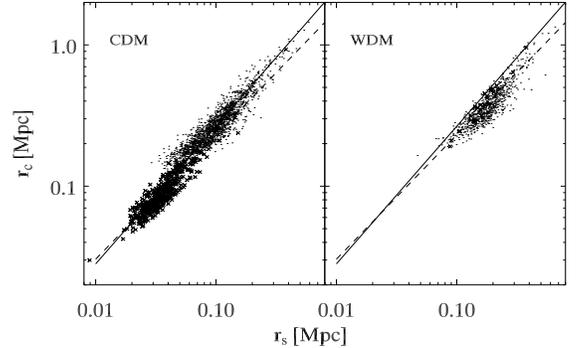}
\figcaption
{Comparison of the core radius, $r_c$, of the Hernquist fit (equation
[\ref{eqn:hern}]) plotted against the scale radius, $r_s$, of the NFW
fit (equation [\ref{eqn:nfw}]) from our CDM and WDM simulations.  Dots
represent halos from the large box run, and crosses are from the small box
run.  The black and dashed lines show the linear fits for the CDM and WDM
cosmologies.  
\label{fig:rcrs}
}
\end{figure}

As with the MAHs, there is little difference between the density
profiles of the WDM and CDM cosmologies for halos with $\mvir \gg
M_c$.  Specifically, for the plotted profiles in this range, most
differences are at the $\sim 1\%$ level.  Significant differences
appear as we near the truncation scale, $M_c$, but, remarkably, the
profiles are still well fit by both the NFW and Hernquist profiles, as
indicated by the residuals in the bottom panels of Figure
\ref{fig:Rho}. We define the concentration in the usual way, $c_{200}
= r_{200}/r_s$.  As noted above, the parameter $r_s$ is
measured by fitting an NFW profile to our profiles using
logarithmically spaced radial bins in the range $0.05\rvir - \rvir$
(similar to \citealp{Bullock01}).  For equilibrated halos at $a = 100$,
the steepness of the halo density profile causes $r_s$ to change
substantially if we fit to a different radius.  By increasing the
outer fit radius to $r_{vir} \approx 2.5\rvir$ as in Bullock \etal
(2001), concentrations typically decrease by a factor of 2 or greater,
depending on the mass of the halo.  A more robust concentration could
alternatively be measured as $c_{halo} = r_{halo} / r_c$ from equation
(\ref{eqn:hern}), although the definition using the NFW fit is more
standard (and our new definition is not well defined for $a \lsim 3$).
Figure \ref{fig:rcrs} plots the relation between the parameters $r_c$ and
$r_s$.  The relation is well fit by a power law, $r_c \propto
r_s^{\nu}$, with $\nu = 0.974 \pm 0.004, 0.88 \pm 0.02$ and
proportionality constants 0.40 and 0.24 for the CDM and WDM
cosmologies.  For what follows, however, we will continue to use the
concentration defined with an NFW fit.

\begin{figure*}
\plotone{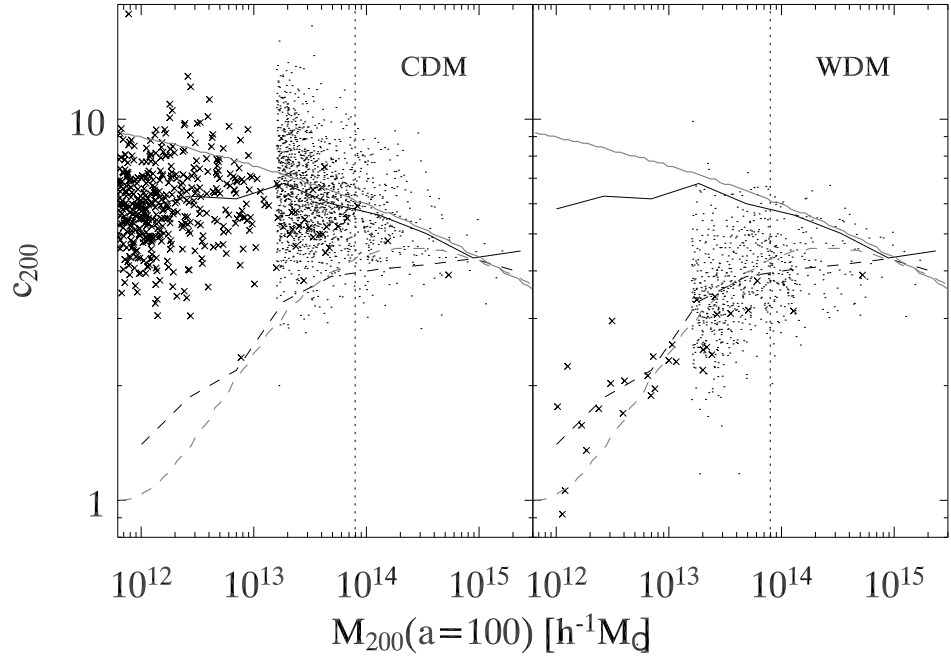}
\figcaption
{The concentration ($c_{200} = r_s / \rvir$)
as a function of mass for CDM and WDM halos.  Dots represent halos
from the large box run, and crosses are from the small box run.  The
dark solid and dashed
lines show average $c_{200}(M_{200})$ values for CDM and WDM halos, and
the vertical dotted line marks the truncation scale, $M_c$.  The light
solid and dashed lines show the calculated $c_{200}(M_{200})$ relation
from the toy model discussed in \S 4.3.  
\label{fig:cofM}
}
\end{figure*}

\begin{figure*}
\plotone{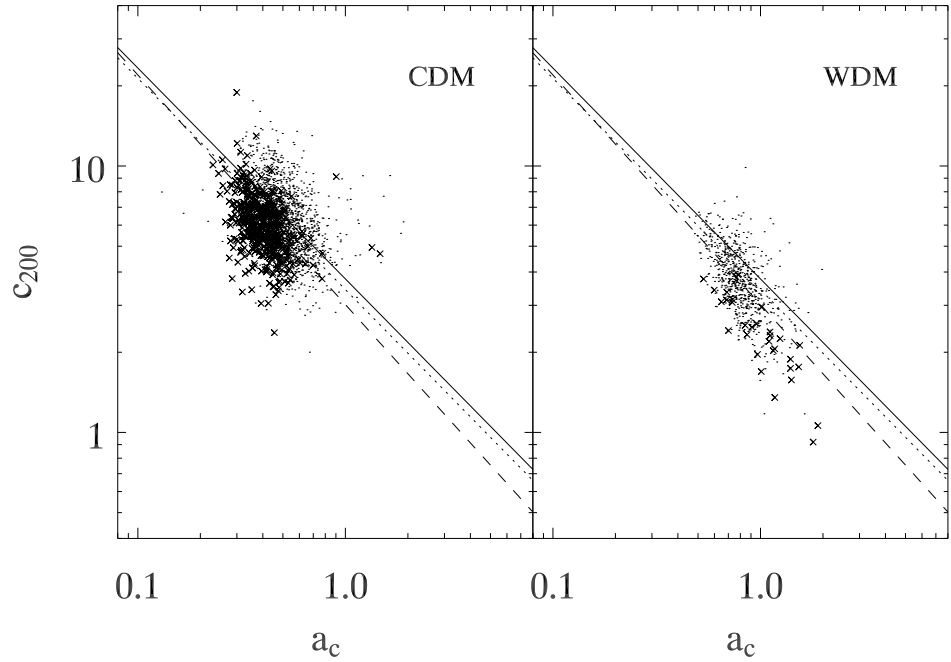}
\figcaption
{The concentration as a function of
formation epoch, $a_c$ from equation (\ref{eqn:Mofa}) for CDM and WDM
halos.  Dots represent halos from the large box run, and crosses are
from the small box run.  The solid line is the trend-line for CDM
halos, and the dashed line is the trend-line for WDM halos.  The dotted
line is the trend-line for the combined sample of all CDM and WDM
halos.  
\label{fig:cofac}
}
\end{figure*}

Figure \ref{fig:cofM} shows the dependence of the concentration
on mass for our halos.  As with Figure \ref{fig:acofM}, the left
panel shows CDM halos and the right panel WDM halos, with dots
representing halos from the large volume realizations and crosses halos
from the small volume realizations.  This plot follows many of the
overall trends of Figure \ref{fig:acofM}.  Again, for $\mvir \gg M_c$
the average values for $c_{200}$ differ by less than 10\% for CDM and
WDM halos and systematically diverge for lower masses.  The low mass
WDM halos are much ``puffier'' than similar mass halos (or their
counterparts) in the CDM simulations.  W02
explained the trend of changing $c_{200}$ with $\mvir$ by identifying
a relationship between $c_{200}$ and $a_c$, which we have reproduced
with our data in Figure \ref{fig:cofac}.  Both the CDM (left panel)
and WDM (right panel) halos follow a power law relation $c_{200}
\propto a_{c}^{-\beta}$, although there is a substantial amount of scatter
present in the CDM relation.  For CDM halos, $\beta = 0.79 \pm 0.2$
(solid line), while the WDM halos have $\beta = 0.86 \pm 0.3$ (dashed
line), values that are both within
$2\sigma$ from the combined slope $\beta = 0.79 \pm 0.1$ (dotted
line).  This finding is similar to the $c \propto a_c^{-1}$ relation
proposed by W02.  Note that for W02, $\rvir$ is replaced with $r_{vir}
\approx 2.5\rvir$ at a =
100 \citep{Eke96}.  The explanation for this trend is that $c_{200}$
is a reflection of the average density at the time of collapse, so
that halos forming earlier should have higher concentrations, exactly
as observed.  This trend holds even for WDM halos with $\mvir \ll
M_c$.  This picture is further enforced by a toy model for
concentrations proposed by \citet{Eke01}, modified to use our density
threshold.  Noting that in WDM models $c_{200}$ decreases with mass
below the truncation scale, Eke \etal (2001) defined an effective
perturbation spectrum,
\be
\sigma_{eff}(M) = -{{\rm d} \sigma(M) \over {\rm d} \ln(M)},
\ee 
so that the spectrum also decreases at low masses.  From this
equation, a collapse epoch, $\actoy$ can be identified as the epoch
where $D(\actoy)\sigma_{eff}(M_s) = 1 /C_{\sigma}$, where $M_s =
M(<2.17r_s)$  (the mass contained within the radius where the circular
velocity of a NFW profile reaches its maximum) and $C_{\sigma}$ is a
fitting parameter.  A central density is defined as in
\citet{Bullock01} such that $\mvir = 4/3\pi r_s^3\tilde{\rho_s}$.
We then assume that 
\be
\tilde{\rho_s} = 200\rho_{crit}(a_0)c_{200}^3
\ee
and set this density scale equal to our overdensity at the epoch of
formation, $200\rho_{crit}(\actoy)$.  This procedure yields the
relation
\be
c_{200} = \left[{\rho_{crit}(\actoy) \over \rho_{crit}(a_0)}\right]^{1/3}.
\ee
The only free parameter in this model is the constant $C_{\sigma}$,
which we set equal to 32.  The results of this model are plotted as
the lighter curves in Figure \ref{fig:acofM}.  The model
characterizes the simulated halos for both the CDM and WDM cosmologies
at $a = 100$, and produces equally good agreement at earlier epochs.
Our toy model differs from that of Eke \etal (2001) only in that we
define our halos in terms of $200\rho_{crit}$, as opposed to the
epoch-dependent quantity $\Delta(a)\rho_{crit}$.  We find that using
$\Delta(a)\rho_{crit}$ requires different values for $C_{\sig}$ at $a
= 1$ and 100.

In all of our simulations, the density
profile is well characterized by an NFW profile for much of the halo's
radial extent, and that the quality of the fit does not
depend on whether a halo was taken from a CDM or WDM cosmology.
The primary difference between halos in these two models --- the
change in concentration parameter --- follows a simple relationship
with the formation epoch that is relatively insensitive to how the
halos form:  CDM halos that have accreted their mass as virialized
clumps; WDM halos with $\mvir > M_c$ that have formed through a period
of rapid smooth accretion followed by growth through accretion of
clumps; and WDM halos with $\mvir < M_c$ that have formed through a
single period of rapid smooth accretion. 

\section{CONCLUSIONS}

This work examines the effects of small scale structure and merger
activity on the formation and ultimate structure of halos in a
$\Lambda$-dominated universe.  Using N-body simulations with an
initial power spectrum that is truncated on small scales, we model a
WDM-like universe where the formation of early low-mass objects is
suppressed and compare the results to standard \LCDM simulations.
Without these seeds for hierarchical growth, halos above
the truncation scale form through an initial rapid accretion phase,
resulting in objects with mass $\mvir \sim M_c =
8\times10^{13}\hinv\msol$.  These objects can then grow through
mergers with other large halos.  Many halos also form below the
truncation scale through incomplete  collapse of
larger perturbations.  These halos form solely through a
monolithic-like collapse process.  Regardless of their size, WDM halos
typically form later and faster, with a larger $\dot{M}_{200}(a)$ than
their counterparts in an \LCDM cosmology.  

To describe this rapid accretion, we generalize the \citet{W02} mass
accretion history formula by introducing a rate index, $\gamma$, that
controls the growth rate evolution, $d\ln(M)/d\ln(a) \propto
a^{-\gamma}$.  This parameter is necessary in order to fit the halo
MAH into the far future.  Otherwise, the halo mass approaches its
asymptote much too slowly. Additionally, for ensemble-averaged halo
histories fit to the present epoch ($a \le 1$), we recover $\gamma
\sim 1.6$, substantially higher than the $\gamma = 1$ assumed in
\citet{W02}. The larger $\gamma$ improves the mean MAH fit by a factor
of 5.  Our generalization also reduces the number of objects that have
unphysical formations epochs, $a_c \gg 1$.  

We have also calculated abundances and decay rates for substructure in
our dark matter halos.  As expected, WDM halos contain much less
substructure, and most host halos destroy all of their
subhalos by the end of our simulations at $a = 100$.  The decay of
subhalos is (on average) exponential in time with half-lives 21Gyr and
7.4Gyr for the CDM and WDM cosmologies.  This is consistent with the
picture discussed below where later formation times create lower
concentration objects that are more prone to disruption.  

Despite differences in the formation process and substructure
abundance, WDM halos exhibit NFW density profiles, just like CDM
halos, albeit with a different $c_{200}(\mvir)$ relationship.  Halos in
WDM cosmologies have lower concentrations than their CDM counterparts,
but follow the same $c_{200}(a_c)$ relation, allowing us to relate
concentration to mass in both cosmologies using a 1 parameter toy model
that characterizes the concentration using the linear power spectrum.
This characterization is motivated by the idea that the concentration
is set by the cosmological background density at the epoch of
collapse.  The form of the halo density profile persists, even though
the method of formation and amount of substructure is changed
substantially in our WDM cosmology.  Taken together, our results
suggest that the form of the halo density profile is set not by
merger activity, but instead through large-scale modes of the
gravitational relaxation process. Halos appear to be very efficient at
erasing their initial conditions and do not care whether their mass
was accreted rapidly, slowly, in clumps, or continuously.  The only
aspect of the density profile directly linked to the halo formation
process appears to be the concentration, which set by the formation
epoch, the epoch when the mass accretion rate drops to a specific
value.  In particular, substructures and major mergers seem to have
little effect in driving the equilibrium structure of a halo. 

The authors would like to thank Volker Springel for the generous use
of his \subfind routine and for useful discussion.  This work has been
supported at the University of Michigan by the Michigan Center for
Theoretical Physics, by NASA grants NAG5-13378 and NNG04GK56G0, NFS
ITR grant ACI-0121671, and by the Foundational Questions Institute via Grant
RFP1-06-1.  M.~T.~B.~was supported by a Predoctoral Fellowship from
the Rackham Graduate School at the University of Michigan.
A.~E.~E.~acknowledges support from the Mill Foundation for Basic
Research in Science at University of California, Berkeley.  

\appendix
\section{Mass Function}

\begin{figure}
\plotone{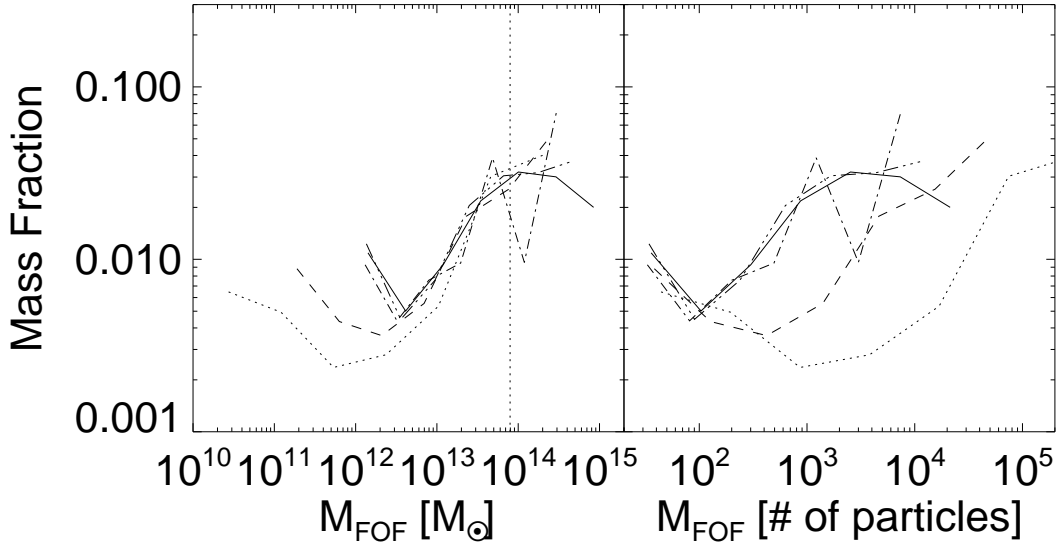}
\figcaption
{{\it Left Panel: } The fraction of mass in a collapsed structure of a
given mass as a function of mass for a number of WDM simulations with
varying degrees of mass particle resolution ($M_{particle} =
6.20\times10^{8} - 3.97 \times10^{10}\hinv\msol$) and simulated volume
($50 - 200 \hinv\mpc$). The vertical dotted line represents the
truncation scale, $M_c$.  {\it Right Panel: } The same mass fractions,
now plotted mass in units of number of particles.
\label{fig:upturn}
}
\end{figure}

The upturn in the FOF mass function (Figure \ref{fig:MassFunction}) at
small mass is a surprising and potentially troubling feature of our
WDM simulations, and it is necessary to determine if this phenomenon is a
result of the simulation or of the FOF group-finding algorithm.  
We have conducted a series of 5 WDM simulations with various volumes
and mass resolutions to examine this effect, and present the resultant
mass fractions at $a = 1$ in Figure \ref{fig:upturn}.  Here, we are
plotting the mass fractions, as opposed to the mass functions shown in
Figure \ref{fig:MassFunction}.  The mass fraction and mass function
are related by $dn(M)/d\ln(M) = (\bar{\rho}/M)f(M)$.  We used a mass
resolution of $3.97\times10^{10} \hinv\msol$ as our base model,
conducting three runs at this scale with comoving box lengths of 200,
100, and 50 $\hinv\mpc$ (solid, dash-dot, and dash-dot-dot-dot lines),
and two additional runs with box length $50 \hinv\mpc$ with mass
resolutions $6.20\times10^{8}$ and $4.96 \times 10^{9} \hinv\msol$
(dotted and dashed lines).  The left panel shows the mass fraction as
a function of mass in units of $\hinv\msol$, while the right panel
shows the same fractions in units of number of particles. 

The most immediate feature of the left panel is that the location
of the upturn changes with mass resolution, independent
of the simulated volume.  The right panel of Figure
\ref{fig:upturn} further indicates that this upturn is a purely
numerical artifact.  Here, the mass fraction is plotted in
units of number of particles in the halo, and we see that at low
masses all resolutions converge on the upturn.  Finally, we also note
that for the two runs that have corresponding CDM cosmologies (solid
and dotted lines) the location of the minimum in the mass fraction
($M_{min}$) corresponds to a transition point: halos with $M >
M_{min}$ have corresponding halos in the CDM simulation, while those
with $M < M_{min}$ do not (see Figure \ref{fig:CrossHalos}).  In
these, 90\% of all halos with $M_{FOF} \gg M_{min}$ have corresponding
halos in the CDM cosmology, while less than 1\% of halos with
$M_{FOF} < M_{min}$ have identified cross halos.   Even though the
input power spectrum for the CDM and WDM cosmologies is vastly
different, it is not the value of $M_c$, but rather the of value
$M_{min}$ that determines the mass where we are unable to find
corresponding halos in the CDM cosmology.  This trend is especially
noteworthy in the case of the small volume realization, where
$M_{min}$ is almost two orders of magnitude below $M_c$.  Finally, the
upturn appears substantially less pronounced for the SO mass
function.  Most of the FOF halos below the upturn have no central
overdensity greater than $200\rho_{crit}$ but instead exist as a more
linear string of particles that are not even identified as
energetically bound by \subfind.  

One expects an upturn eventually, since a FOF group finder puts every
particle in a group of some size.  At any given time, 50-70\% of all
particles end up in ``groups'' of one particle, requiring an upturn
somewhere, something that you would expect even in a standard
\LCDM simulation (although the effect there should be much smaller due
to the large number of actual groups at this mass range).  Our results
suggest that this contamination is an issue for groups containing
fewer than
\be
N_{min} \approx 250 \left({M_{particle} \over 10^{10} \msol
\hinv}\right)^{-0.56}
\ee 
particles when a linking length of 0.164 is used.

\end{document}